\documentclass[a4paper,10pt]{article}
\usepackage[utf8]{inputenc}
\usepackage{graphicx}
\usepackage{color}
\begin{document}

\def\pa{\partial}
\def\vb{{\bf v}}
\def\Fb{{\bf F}}
\def\Kb{{\bf K}}
\def\Bb{{\bf B}}
\def\ol{\overline}
\def\eb{{\bf e}}
\def\omb{{\bf \omega}}
\def\MC{meridional circulation}
\def\DF{differential rotation}
\def\CZ{convection zone}

\title{The Meridional Circulation of the Sun: Observations, Theory and Connections with the Solar Dynamo}         
\author{Arnab Rai Choudhuri \\ Department of Physics \\ Indian Institute of Science 
\\ Bangalore - 560012, India}        
\date{}          
\maketitle

\begin{abstract}
The meridional circulation of the Sun, which is observed to be poleward at the surface,
should have a return flow at some depth.  Since large-scale flows like
the \DF\ and the \MC\ are driven by turbulent stresses in the convection zone, these
flows are expected to remain confined within this zone.  Current observational (based 
on helioseismology) and theoretical (based on dynamo theory) evidences point towards
an equatorward return flow of the meridional circulation
at the bottom of the convection zone.  Assuming the
mean values of various quantities averaged over turbulence to be axisymmetric,
we study the large-scale flows in solar-like
stars on the basis of a 2D mean field theory. Turbulent stresses in a 
rotating star can transport angular momentum, setting up a differential rotation. 
The meridional circulation arises from a slight imbalance between two
terms which try to drive it in opposite directions: a thermal wind term (arising
out of the higher efficiency of convective heat transport in the polar regions) and a
centrifugal term (arising out of the \DF). To make these terms comparable, the poles
of the Sun should be slightly hotter than the equator. We discuss the important role
played by the \MC\ in the flux transport dynamo model.  The poloidal field generated
by the Babcock--Leighton process at the surface is advected poleward, 
whereas the toroidal field produced at the bottom of
the convection zone is advected equatorward. The fluctuations in
the \MC\ (with coherence time of about 30--40 yr) help in explaining many aspects
of the irregularities in the solar cycle.  Finally, we discuss how the Lorentz force
of the dynamo-generated magnetic field can cause periodic variations in the large-scale
flows with the solar cycle.

\bigskip

\noindent{\bf Astrophysical plasma, Solar physics, Sun: solar magnetism, Sun:  helioseismology,
Fluid flow: rotational}

\medskip

\noindent{\bf PACS Number(s)}: 95.30.Qd, 96.60.-j, 96.60.Hv, 96.60.Ly, 47.32.-y 

\end{abstract}

\section{Introduction}       

There is an intriguing fluid flow pattern inside the Sun (and probably inside other
solar-like stars): the {\em meridional circulation}. It is known for nearly half a century
that matter at the solar surface moves continuously from the equator to the poles in
both the hemispheres---the maximum speed of this motion at mid-latitudes being of order
20 m s$^{-1}$.  Since we do not expect matter to be piled up near the poles, there has to
be a return flow at some depth underneath the Sun's surface bringing back the matter
from the polar regions to the equatorial region.  Apart from the intrinsic interest we
may have in a such a flow from a purely fluid dynamical point of view, it is realized 
in the last few years that this flow plays a crucial role in the dynamo process producing
the 11-year sunspot cycle.  Let us begin with a discussion of the mathematical definition
of the meridional circulation.

Any plane passing through the rotation axis of a rotating, self-gravitating body (such as a star 
or a planet) is referred to as a {\em meridional plane}.  If we introduce spherical coordinates   with 
the origin at the centre of the body and with the rotation axis as the polar axis, it is easy to 
see that a meridional plane would be an $(r, \theta)$ plane over which $\phi$ is constant.  Let 
us begin by considering a simple kind of fluid flow which is axisymmetric (i.e.\ independent of 
$\phi$).  The fluid velocity can be written as
$$\vb = v_r (r, \theta, t) \, \eb_r + v_{\theta} (r, \theta, t) \, \eb_{\theta} + 
v_{\phi} (r, \theta, t) \, \eb_{\phi}.\eqno(1)$$
The part $v_{\phi} (r, \theta, t) \, \eb_{\phi}$ is called the {\em azimuthal} or {\em zonal circulation}, 
whereas the part lying in the meridional plane, i.e.
$$\vb_m = v_r (r, \theta, t) \, \eb_r + v_{\theta} (r, \theta, t) \, \eb_{\theta} \eqno(2)$$ 
is called the meridional circulation. Writing $v_{\phi} = r \sin \theta \: \Omega$, where
$\Omega$ is the angular velocity, we can put (1) in the form
$$\vb = \vb_m + 
r \, \sin \theta \: \Omega (r, \theta, t) \, \eb_{\phi}.\eqno(3)$$
Very often we consider flows which are time-independent in addition to being
axisymmetric. It then easily follows from the equation of continuity that 
$\nabla. (\rho \vb_m) = 0$, 
implying that flows in the meridional plane should be of the nature of circulation with
closed streamlines. 

We are aware of fluid flows existing in the interiors and atmospheres of many stars 
and planets. A state of strict hydrostatic equilibrium without any motions is often unstable
or is continuously disturbed by forces driving the flows. The flows
inside stars and planets are sometimes of the nature of turbulent flows, which means that they 
are neither axisymmetric nor time-independent.  However, by suitable spatial and temporal 
averaging, we can often get a mean flow pattern which may be approximated as axisymmetric 
and time-independent---at least over a certain regime of space and time.  Considering such flows 
is the natural first step in understanding the complex physics of this subject. In the theoretical
portions of this basic review, we shall always restrict our discussion to mean \MC s which are 
axisymmetric, but we shall discuss certain aspects of time variations. 

Before getting into a discussion of the meridional circulation of the Sun, let us consider a 
simple meridional circulation in the Earth's atmosphere.  Suppose the equatorial region 
of the atmosphere is heated by the Sun's rays.  The air there expands and becomes lighter, 
causing it to be buoyant and to rise up. The colder air from higher latitudes would rush 
to the equatorial region.  The hot air, which rises in the equatorial region, will cool 
as it rises and then will flow to the higher latitudes through the upper layers of the 
atmosphere, thereby setting up a meridional circulation pattern.  At first sight, 
it may seem that the physics of this problem is straightforward.  After all, it involves 
only thermodynamics and fluid mechanics.  We invite those readers who are familiar with 
thermodynamics and fluid mechanics, but have not studied the theory of meridional circulation 
earlier, to set up the mathematical equations of this problem.  As soon as we try to make a 
mathematical formulation of this problem, we realize that it is much more complicated than 
what we may initially think. 

The best way of handling such a fluid flow problem is to consider the {\em vorticity}   
$$\omb = \nabla \times \vb.\eqno(4)$$ 
It is easy to see that the meridional circulation given by (2) would produce a $\phi$ component of 
vorticity.  One can try to obtain an equation for $\omega_{\phi}$ from the basic equations of 
fluid mechanics.  Usually the equation for $\omega_{\phi}$ turns out to have the form
$$\frac{\pa \omega_{\phi}}{\pa t} = ({\rm source \; terms}) + ({\rm dissipation \; term}). \eqno(5)$$
As we shall see later, the meridional circulation of the Sun really satisfies an equation like 
this. The source terms, which may involve thermodynamic considerations, drive the meridional 
circulation, whereas the dissipative term tries to damp it. If these terms somehow manage to 
balance each other, then we may get a time-independent \MC. 

At the outset, let us point out an important result of stellar structure modelling that the
heat generated by nuclear reactions at the centre of the Sun is transported outward by radiative
transfer till about $r = 0.7 R_{\odot}$ (where $R_{\odot}$ is the solar radius), whereas heat is
transported by convection from $r = 0.7 R_{\odot}$ to $r = R_{\odot}$ [1, 2].  In other words, we
have a turbulent convection zone just below the Sun's surface. The convection cells at the
solar surface known as {\em granules} can be observed through telescopes.
As we shall discuss later, the turbulent
stresses in the convection zone play a crucial role in driving the \MC.  So, it is assumed 
that the streamlines of the \MC\ would remain confined within the convection zone. While
developing the theory of the \MC\ of the Sun, we shall see that this theory is intimately connected
with the theory of differential rotation (a non-constant $\Omega$ varying with $r$ and $\theta$
is called {\em differential rotation}). It has been known from the mid-nineteenth century that the
angular velocity at the solar surface near the equator is more than that at higher latitudes [3].  As we
shall point out later, the new science of helioseismology has provided the crucial information of how
the angular velocity $\Omega (r, \theta)$ varies under the solar surface.  Helioseismology
also provides information about the \MC\ underneath the solar surface.  However, as we shall
discuss, this information becomes less and less reliable as we go deeper down from the solar
surface, and although there is now strong observational evidence that the return
flow of the \MC\ (bringing back matter from the polar regions to the equatorial region) takes
place at the bottom of the convection zone, there is still not a complete consensus
on this. A poleward flow at the solar
surface and an equatorward flow deeper down give rise to negative $\omega_{\phi}$ within the
core region of the \MC\ in the northern hemisphere.  Hence, when we develop the theory of the solar
meridional circulation, the sum of the source terms in (5) is expected to be negative in
much of the northern hemisphere.

Sunspots are regions of concentrated magnetic field (typically of order 3000 G) and the 11-year
sunspot cycle (also called the {\em solar cycle}) 
is the magnetic cycle of the Sun.  This cycle is believed to be caused by
a magnetohydrodynamic or MHD process known as the {\em dynamo process}.  The early models of the
solar dynamo were developed at a time when the existence of the \MC\ was not known and these
early models naturally did not include the \MC.  Over the years, it became clear that these
earlier models of the solar dynamo without the \MC\ had many difficulties.  From the 1990s,
a new kind model known as the {\em flux transport dynamo model}---in which the \MC\ plays
a crucial role---has been developed. This model has been successful in explaining various
aspects of the solar cycle, leading to an increased interest in the science of 
the \MC.   

It may be mentioned that, in
the last few years, there have been some impressive numerical simulations of
convection inside rotating stars,
showing that the turbulent stresses can drive the large-scale flow patterns.  
The discussion of simulations will be rather limited in this review, the focus being
on the mean field theory obtained by averaging over turbulence---for the following two
reasons.  Firstly, the primary aim of this review is to elucidate 
the basic physics, which can be understood
better from the mean field theory rather than from a description of the results of
simulations.  Secondly, the author personally is not particularly qualified to discuss the intricacies
and subtleties of numerical simulations.  We shall highlight some results of simulations
which throw light on our discussions based on the mean field model, but we shall not
attempt to present any systematic account of the simulations done by different groups. 

We shall summarize the salient features of the observational data about the \MC\ of the Sun
in the next Section.  Then \S3 will be devoted to discussing the basic theory of the \MC---along
with the basic theory of differential rotation---presenting some modelling efforts.  The role
of the \MC\ and its irregularities in the flux transport dynamo model of the Sun will be discussed
in \S4.  Then \S5 will be devoted to the back reaction of the dynamo on the large-scale flow
patterns of the Sun.  Finally, we shall present some concluding remarks in \S6.    

\section{Relevant observational data}

One of the challenges of observing the \MC\ at the solar surface is that it involves fluid
flows which are much weaker than other kinds of fluid flows present there.  We have mentioned
that the maximum speed of the \MC\ at the mid-latitudes is about 20 m s$^{-1}$, which means
that the time to traverse a quadrant of the Sun's circumference (from the equator to the pole)
would be of the order of about 1.7 yr.  The solar surface has other kinds of fluid flows which
are much faster with shorter time scales.  The convective velocities associated with granules
at the solar surface are of order 1--2 km s$^{-1}$, the typical lifetimes of granules being of
the order of a few minutes.  The rotation period of about 25 days near the solar equator gives
rise to an azimuthal velocity of about 2 km s$^{-1}$. Thus, to measure the velocity of the \MC\
directly, we have to pick up a signal much weaker than the other signals present. 

\subsection{Meridional circulation at the solar surface}

Although the \MC\ is much weaker than other fluid flows at the solar surface, it can be identified
by something special that it does.  It carries various surface features with it poleward.  We may
mention that the turbulent velocities of convection near the solar surface 
make things spread out, giving rise to an
effective diffusion, which is very important in the flux transport dynamo model.  This diffusion
also can cause a poleward spread of various things.  However, the role of the \MC\ in the poleward
transport of surface features is much more direct and effective (the spread by diffusion goes as $\sqrt{t}$,
whereas the transport by a flow goes as $t$). Historically, the \MC\ was discovered from
observations of such poleward transport. 

\begin{figure}
\centerline{\includegraphics[width=0.8\textwidth,clip=]{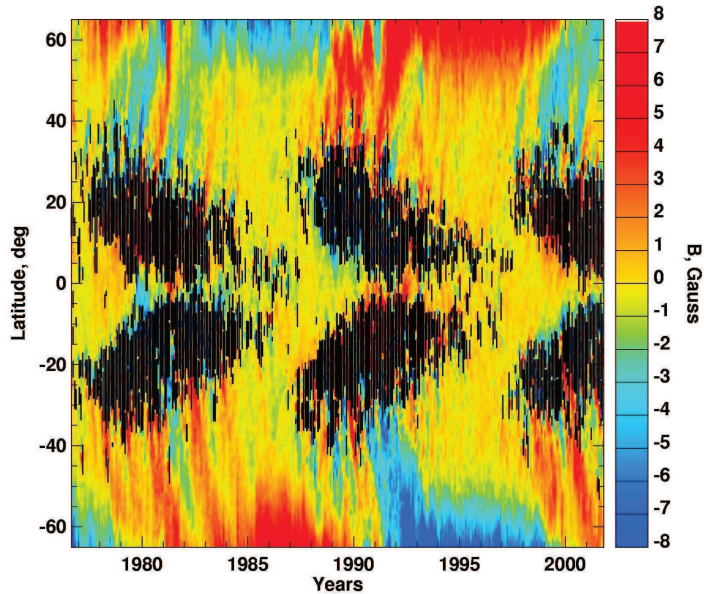}}
\caption{A time-latitude plot in which the shaded regions indicate the solar latitudes at
which sunspots were seen as different times. The colours indicate values of the 
longitude-averaged magnetic field outside the sunspots.}
\end{figure}

Sunspots, regions of strong magnetic field on the solar surface, appear around latitudes 30$^{\circ}$--40$^{\circ}$ at
the beginning of a solar cycle.  As the cycle progresses, sunspots appear at lower and lower
latitudes.  The shaded portions in Figure~1 indicate the regions in a time-latitude plot where
sunspots appeared during the span of a little more than 
two solar cycles.  The colours in Figure~1 indicate the longitude-averaged
values of the magnetic field outside sunspots in this time-latitude plot.  The magnetic field outside 
sunspots consists of latitude belts in which this field is predominantly of a particular sign.  In contrast to the
sunspot belts which shift equatorward with the solar cycle, the field outside sunspots seems to
be advected poleward, suggesting a poleward flow of matter which carries this magnetic field with
it.  The existence of the \MC\ was first inferred from the observation that there were 
unipolar patches of magnetic field in certain latitude belts [4] and that these patches shifted poleward
with time [5, 6].  The polar field of the Sun, which gets built up as the magnetic fields from the lower
latitudes are brought to the polar region, reverses its direction around the time of the sunspot
maximum and is clearly tied to the solar cycle.  The early measurements with low-resolution 
magnetograms suggested that the polar field is of order 10 G (see the colour code in Figure~1).
We know for several decades now that this magnetic field outside sunspots is actually concentrated 
inside highly intermittent flux
tubes with magnetic field of order 1000 G [7] and the values measured by the early low-resolution
magnetograms are merely values averaged over patches of the solar surface when the flux tubes are
not resolved. The \MC\ could be estimated also from the poleward displacements of small
magnetic features [8].

Apart from the poleward shift of unipolar magnetic patches, there is another important proxy which
gave a lot of information about the \MC\ in the early years of research in this field.  There must
be a neutral boundary line between the regions of opposite magnetic polarity on the solar surface.
When we observe the Sun using an H$_{\alpha}$ filter, we often see dark filaments above the neutral
line---presumably made out of cool gas resting on the magnetic canopy that must exist above the
neutral line.  Positions of the dark filaments in an H$_{\alpha}$ plate would indicate the neutral line
and one can draw inferences about the \MC\ from a study of how the neutral line shifts poleward with
time.  From an analysis of the H$_{\alpha}$ plates of the Sun taken at the Kodaikanal Observatory
over several decades, the existence of the \MC\ in the early decades of the twentieth century, when
there were no measurements of the magnetic field outside sunspots, could be established [9, 10]. 

Some attempts to measure the \MC\ at the surface directly through the Doppler shifts of spectral
lines have also been made [11, 12].

\subsection{Sub-surface results from helioseismology}

After the existence of the \MC\ at the solar surface was established, the important question was
whether we can determine its nature underneath the surface---especially whether we can find where
the return flow from the poles to the equator occurs.  During the last few years, we have some information
about it from helioseismology, which is the study of solar oscillations first discovered at the
solar surface in the 1960s [13].  These surface oscillations are caused by acoustic waves propagating
underneath the solar surface and buffeting the surface.  If there are large-scale fluid flows underneath
the surface, they affect the propagation of the acoustic waves and it is possible to draw inferences
about these flows from the analysis of the oscillations data.  This is a highly technical subject and
the details of how this is done are beyond the scope of this review.  We refer the interested readers
to standard reviews of this subject [14, 15, 16] and present only the results here.

\begin{figure}
\centerline{\includegraphics[width=0.85\textwidth,clip=]{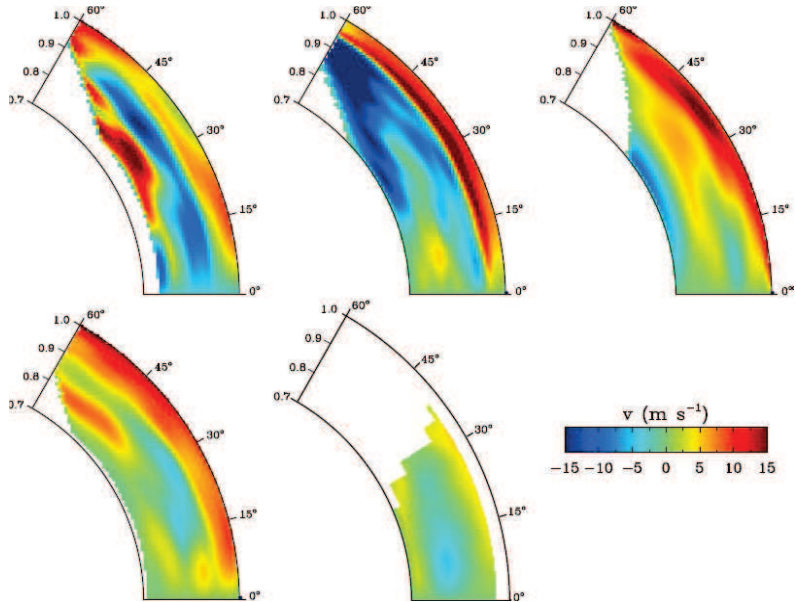}}
\caption{Hemispherically symmetrized profiles of the \MC\ component $v_{\theta}$ obtained by
(a) Zhao et al. [19], (b) Jackiewicz et al. [21], (c) Rajaguru
and Antia [20], (d) Chen and Zhao [22], and (e) Lin and Chou [23].
Poleward flows are positive and equatorward flows are negative.
Figure courtesy of Junwei Zhao.
}
\end{figure}

The effect of the \MC\ on the solar oscillations at the surface is a very small effect and it becomes
increasingly difficult to make inferences about the \MC\ in deeper layers of the Sun underneath the
surface from this small effect.  The first results of helioseismology were about the nature of the
\MC\ in the layers immediately underneath the solar surface [17, 18].  Only within the last few years, there
have been serious efforts to look for the return flow from the poles to the equator.  As we shall point
out in our discussion of the flux transport dynamo model, we get the best results if we assume that
there is only one cell of the \MC\ spanning the entire convection zone, with the return flow at the
bottom of the convection zone.  Whether helioseismic studies can either confirm or contradict this view
has become a very important question.  Some authors claim that they find evidence for a return flow
at the middle of the convection zone rather than at the bottom [19], whereas other authors, analyzing the
same data, conclude that a single-cell \MC\ spanning the whole of the convection zone is consistent
with the data [20].  Figure~2 shows results presented by different groups about the nature of the \MC\
underneath the solar surface.  As we can easily see, there are large divergences among the results
of different groups for the \MC\ in the deeper layers of the convection zone.  
A very recent analysis 
of data from different sources led Gizon et al. [24] to conclude that the \MC\ consists of a single cell
in each hemisphere, with the return flow at the bottom of the convection zone.

\subsection{The differential rotation of the Sun}

As we have pointed out in the Introduction and shall again see in \S3, the theory of the \MC\
is intimately connected with the theory of the other large-scale fluid pattern inside the
Sun: the differential rotation given by a non-constant $\Omega (r, \theta)$. Hence, a basic
knowledge about the nature of the differential rotation is essential for our discussion.
One of the remarkable achievements of helioseismology is that it has provided a map of 
$\Omega (r, \theta)$ underneath the solar surface.  The early maps obtained in the 1980s
eventually converged to a robust map by the mid-1990s [25, 26].  Figure~3 is a map of
meridional circulation inside the Sun.

\begin{figure}
\centerline{\includegraphics[width=0.6\textwidth,clip=]{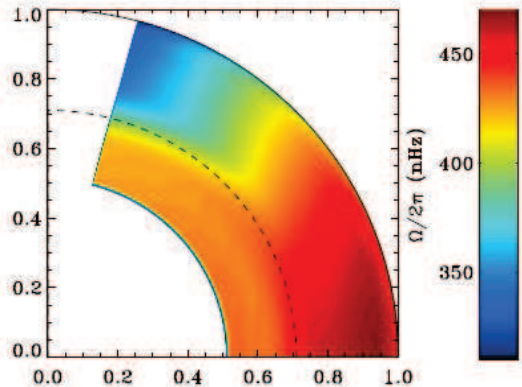}}
\caption{A profile of the differential rotation inside the Sun obtained by helioseismology.
From Howe et al. [27], as presented in Basu [16].}
\end{figure}

We find that the \DF\ is confined within the convection zone, with indications that the
radiative core of the Sun may be rotating like a solid body.  This result is along theoretically
expected lines because we think that the \DF\ also, like the \MC, is driven by turbulent
stresses in the convection zone, as we shall discuss in \S3.2. Within the convection zone,
$\Omega (r, \theta)$ appears to be approximately constant on conical surfaces, with the regions
near the equator having higher values of $\Omega (r, \theta)$ compared to the regions near
the pole.  Such a distribution of $\Omega (r, \theta)$ within the main body of the convection
zone results in a strong radial gradient of $\Omega (r, \theta)$ at the bottom of the \CZ,
which can be seen in Figure~3.  This region of strong gradient of $\Omega (r, \theta)$ at the
bottom of the \CZ\ is called the {\em tachocline}.

We may point out that asteroseismology (i.e.\ the study of stellar oscillations) has now
started giving some results of differential rotation in solar-like stars [28].

\subsection{Variations of the \MC\ with time}

We now come to the important question whether there are variations of the \MC\ with time.
On simple theoretical grounds, we may expect a systematic variation with the solar cycle.
Presumably, the magnetic field in the solar interior is strongest at the time of the sunspot
maximum and the Lorentz force due to this magnetic field also must be strongest.  This
Lorentz force may act on the large-scale flows and may cause a variation with the solar
cycle.  The variation of the \MC\ with the solar cycle has indeed been 
found---both from
helioseismology [29, 30, 31, 32] and from the tracking of surface traces [33]. Figure~4
shows how the \MC\ at a mid-latitude point on the solar surface varied with time during a solar cycle,
with the sunspot number plotted along with it.  It is clear that the \MC\ becomes weaker
at the time of the sunspot maximum. The equatorward \MC\ at the bottom of the convection
zone is also now found to be weaker at the time of the solar maximum [24]. 
We expect the Lorentz force of the solar magnetic field
to act on the \DF\ also.  The variations of the \DF\ with the solar cycle, known as {\em 
torsional oscillations}, have been studied extensively. Although we shall make a few comments
on torsional oscillations in \S5, a detailed discussion of
torsional oscillations is outside the scope of this review (see [34] 
and references therein). We point out another intriguing
aspect of the \MC\ at the sunspot maximum. There seems to be an inward flow towards the
sunspot belt superposed on the overall flow pattern [35, 32].  

\begin{figure}
\centerline{\includegraphics[width=0.6\textwidth,clip=]{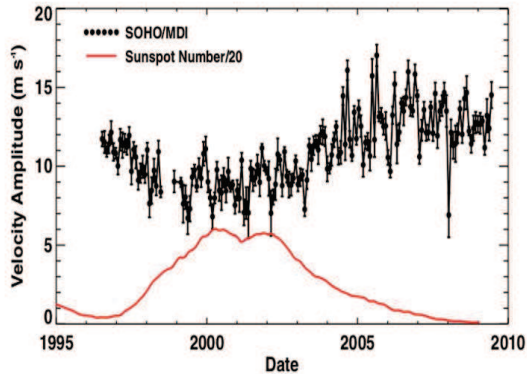}}
\caption{The variation with time of the \MC\ at the surface at mid-latitudes, plotted along
with the sunspot number.  From Hathaway and Rightmire [33]. }
\end{figure}
Apart from the systematic variation with the solar cycle, are there non-systematic random
fluctuations in the \MC?   Since
we have reliable observational data of the \MC\ for a period not longer than a quarter
century, we cannot directly conclude from these data whether there had been fluctuations
in the \MC\ with longer coherence times.  However, we can try to draw some inferences about
this from indirect considerations.  As we shall discuss in \S4.2, the period of the flux
transport dynamo decreases with the amplitude of the \MC.  This means that the solar
cycle durations will be shorter when the \MC\ is stronger and vice versa.  Figure~5 is a plot
of the durations of last 23 solar cycles spanning over more than a 
couple of centuries. There have been
epochs when successive solar cycles had durations shorter than the average, suggesting that
the \MC\ was stronger during such epochs.  From such indirect considerations, we can conclude
that there have been fluctuations in the \MC\ in the past with coherence times of order
30--40 yr [36]. Passos and Lopes [37] reconstructed the history of the \MC\ 
in the past 250 years on the
basis of a low order dynamo model and arrived at very similar conclusions.

\begin{figure}
\centerline{\includegraphics[width=0.9\textwidth,clip=]{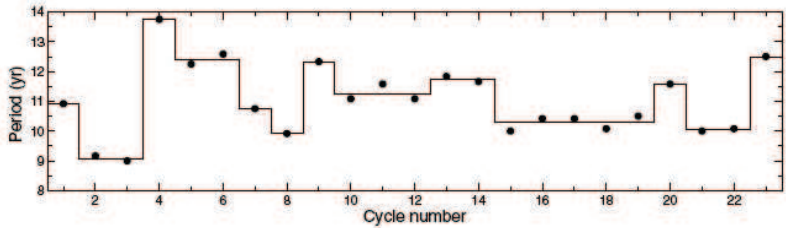}}
\caption{The points show the durations of the last 23 solar cycles against the cycle
number.  The solid line is indicative of the trend in variations of the cycle
durations.  From Karak and Choudhuri [36]. }
\end{figure}

\section{Theory of \MC}

The theory of the \MC\ happens to be a somewhat complicated subject. While the majority
of the research papers on this subject would appear fairly forbidding and
inaccessible to the uninitiated, we are also not aware of any convenient
textbooks or pedagogical reviews from which a beginner can learn this subject.
The two classic monographs by Tassoul [38] and R\"udiger [39] which 
discussed large-scale flows inside stars in some
detail are now very much outdated.  Both these monographs give comprehensive historical
summaries of early research in this field before reliable observational data for
the \MC\ and the internal \DF\ of the Sun became available and before the currently
held theoretical viewpoint emerged. The present review mainly focuses on the
current theoretical viewpoint based on observational data, without much
discussion of the earlier efforts.
We refer the readers to a couple of excellent short reviews by Kitchatinov 
[40, 41], on which we draw heavily in our presentation.
Since the \MC\ now appears to be so important in many solar phenomena, it is
desirable that a solar physicist should have a rough, qualitative idea about the
theory of how the \MC\ arises.  Our aim is to provide that in this Section.
We assume that readers are familiar with the basic principles of fluid
mechanics and MHD (will be needed in the next two sections), which are discussed
in many well-known books.

\subsection{Governing equations for large-scale fluid motions}

Any discussion of the dynamics of fluid flows should begin with the
Navier--Stokes equation, which we write in the following form
$$\frac{\pa}{\pa t}(\rho \, v_i) + \frac{\pa}{\pa x_j}(\rho \, v_i v_j) = 
- \, \frac{\pa p}{\pa x_i} + \rho F_i + \frac{\pa}{\pa x_j}\left( 
\mu \frac{\pa v_i}{\pa x_j} \right). \eqno(6)$$
(See for example [42], \S7, \S15; [43], \S4.3, \S5.1). 
Here all the symbols have their usual meanings, ${\bf F}$
being the body force per unit mass (like gravity).  We are also using the
summation convention that an index repeated twice implies summation over
the spatial directions. We shall restrict our discussion to unmagnetized
fluids in this Section, with discussions about the magnetic field postponed
to the next two Sections.

When we deal with turbulent fluid motions (as within the convection zone
of the Sun), we can write the velocity in the following manner
$$v_i = \, \ol{v_i} + v_i', \eqno(7)$$
where $\ol{v_i}$ is the mean value of $v_i$ averaged over turbulence and
$v_i'$ is the fluctuation around the mean. Let us now write $v_i$ and $v_j$
in (6) in this manner and average over turbulence.  Keeping in mind that
$\ol{v_i'} = 0$, we are led to
$$\frac{\pa}{\pa t}(\rho \, \ol{v_i}) + \frac{\pa}{\pa x_j}(\rho \, \ol{v_i}\;
\ol{v_j} + \rho \; \ol{v_i' v_j'}) = - \,
\frac{\pa p}{\pa x_i} + \rho \, F_i + \frac{\pa}{\pa x_j}\left(\mu \, \frac{\pa 
\ol{v_i}}{\pa x_j} \right). \eqno(8)$$
Subtracting the equation of continuity
$$\frac{\pa \rho}{\pa t} + \frac{\pa}{\pa x_j} (\rho \, \ol{v_j}) = 0$$
multiplied by $\ol{v_i}$ from the left side of (8), we get
$$\rho \frac{\pa \ol{\vb}}{\pa t} + \rho \, (\ol{\vb}. \nabla) \ol{\vb} = -
\nabla p + \rho \, \Fb + \Kb, \eqno(9)$$
where $\Kb$ is a term of which the $i$-th component given by
$$K_i = \, \frac{\pa}{\pa x_j}\left( - \, \rho \,  \ol{v_i' v_j'} + \mu \, \frac{\pa \ol{v_i}}{\pa x_j} \right) \eqno(10) $$
involves the turbulent stress tensor $\rho \, \ol{v_i' v_j'}$, which plays
a crucial role in the theory of large-scale flows inside the Sun. 
Lebedinski [44] appears to be the first person to realize in 1941 that turbulent
stresses may drive large-scale flows.  ``To remind us of his contributions'', the 
driving of mean large-scale flows by turbulent stresses has been christened as
the {\em $\Lambda$-effect} by R\"udiger ([39], p.\ 37).  This idea was further
developed by Wasiutynski [45] and Biermann [46].

Since we shall be primarily dealing with mean fluid flows, let us drop
the overline sign henceforth and write $\ol{\vb}$ as $\vb$, keeping
in mind that from now onwards $\vb$ would refer to the mean flow.  We
write (9) as follows
$$\frac{\pa \vb}{\pa t} + \nabla \left( \frac{1}{2} v^2 \right)
- \vb \times (\nabla \times \vb) = - \,
\frac{\nabla p}{\rho} +  \Fb + \frac{\Kb}{\rho}. \eqno(11)$$
This is going to be the central equation in our theoretical discussions.
Since the turbulent stress term $\rho \, \ol{v_i' v_j'}$ in (10) is
usually several orders of magnitude larger than the viscous stress
term $\mu (\pa v_i/\pa x_j)$ inside a stellar convection zone, often
the viscous stress term is neglected in (10).  However, turbulence
itself gives rise to an effective viscosity and sometimes the turbulent
stress term is taken to be as follows
$$\rho \, \ol{v_i' v_j'} = - \mu_T \, \frac{\pa v_i}{\pa x_j}, \eqno(12)$$
where $\mu_T$ is the turbulent viscosity.  
(Strictly speaking, we should symmetrize any viscous tensor term to
exclude rotation, e.g.\ [43], p.\ 79---we are being somewhat hand-waving
here). It follows
from (10) and (12) that
$$\Kb = \mu_T \, \nabla^2 \vb, \eqno(13)$$
where we have neglected the term due to the viscous stress tensor (arising
out of `molecular' viscosity). However, we should stress that (12) and (13)
are approximations which sometimes miss out some of the essential physics
connected with the large-scale flows inside the Sun.  When we want to
make a realistic model of the large-scale flows inside the Sun, we have
to determine the viscous stress tensor $\rho \, \ol{v_i' v_j'}$ more carefully
to look for non-dissipative parts. Still, in our discussions, we shall sometimes point
out as to what happens if $\Kb$ is given by (13), since this simplification
often gives us quite a bit of insight into the nature of the problem.

In a 1963 pioneering paper, Kippenhahn [47] took the turbulent stress term
to be of the form (13).  However, he assumed the coefficient $\mu_{T,r}$
for the radial transport of momentum to be different from the coefficient
$\mu_{T,h}$ for the horizontal transport of momentum.  This already gave
very interesting results which we shall discuss in \S3.2.  
Durney and Spruit [48] were among the first authors to attempt a
detailed calculation of the turbulent stress tensor for convection
inside a rotating star. Later,
Kitchatinov and R\"udiger [49, 50] calculated this tensor
from 
their model of turbulence and constructed details models of large-scale
flows inside rotating stars.
A look at these papers [48, 49, 50] shows the complexity
of the expressions  of the turbulent stress tensor which these authors arrived at.
We shall try to discuss some of the basic physics of the problem without
getting into the details of how to calculate the turbulent stress tensor.
In \S3.2 we shall indicate how to compute only one crucial component
$\ol {v_r' v_{\theta}'}$ of the turbulent stress tensor.

Let us now consider the $\phi$ component of (11) in spherical 
coordinates.  When we assume
axisymmetry (i.e.\ $\pa/\pa \phi = 0$ everywhere), the gradient terms
do not have any component in the $\phi$ direction and a body force
like gravity would also have no $\phi$ component.  Then we get
$$\frac{\pa v_{\phi}}{\pa t} 
- [\vb \times (\nabla \times \vb)]_{\phi} = \frac{K_{\phi}}{\rho}. 
\eqno(14)$$
This is the basic equation governing the dynamics of the \DF.

As pointed out in the Introduction, we need to find an equation
for $\omega_{\phi}$ of the form (5) in order to develop a theory
of the \MC.  We need to take the curl of (11) and consider its
$\phi$ component.  This gives
$$\frac{\pa \omega_{\phi}}{\pa t} =
[\nabla \times \{ \vb \times (\nabla \times \vb) \} ]_{\phi} -
\frac{1}{\rho^2} \, [\nabla p \times \nabla \rho]_{\phi} 
+ \left[\nabla \times \left(\frac{\Kb}{\rho}
\right) \right]_{\phi}. \eqno(15)$$
This is the basic equation governing the dynamics of the \MC.

As we shall point out in \S3.2 and \S3.3, the terms  
$[\vb \times (\nabla \times \vb)]_{\phi}$ and
$[\nabla \times \{ \vb \times (\nabla \times \vb) \} ]_{\phi}$ 
appearing in (14) and (15) involve both
the \MC\ and the \DF.  Because of these terms, (14) and (15) get coupled
to each other and we cannot solve one of them in isolation.  Both
of them have to be  solved together, showing that the theories of the
\DF\ and the \MC\ are intimately connected to each other. To solve
these equations, we need to know the turbulent stress 
$\rho \, \ol{v_i' v_j'}$ so that we can calculate $\Kb$ by using (10).
So, in order to develop theories of the
\DF\ and the \MC, we need to proceed as follows. We first have
to evaluate the turbulent stresses from some suitable theory of
turbulence.  Then we can solve (14) and (15) together.  Often we may
be interested in the steady large-scale flows in the interior of a
star.  Then the time evolution terms in (14) and (15) can be set
to zero.  Still, it is an immensely difficult problem to solve (14)
and (15).  We discuss some basic physics issues connected with this
problem in the next two subsections.

\subsection{Driving the \DF}

Although this is a review primarily devoted to the \MC, we shall see
in \S3.3 that we need to know the profile of the \DF\ to calculate
the main driving term for the \MC.  So we begin with a discussion
of the theory of \DF.  As we already pointed out, (14) gives the
dynamics of the \DF.  For an axisymmetric velocity field given by
(1), we can easily work out the expression for 
$[\vb \times (\nabla \times \vb)]_{\phi}$ so that (14) leads to
$$\frac{\pa v_{\phi}}{\pa t} + \frac{v_r}{r} \frac{\pa}{\pa r}(r \, v_{\phi})
+ \frac{v_{\theta}}{r \sin \theta} \frac{\pa}{\pa \theta}
(\sin \theta \, v_{\phi}) = \frac{K_{\phi}}{\rho}. 
\eqno(16)$$
The second and third terms in this equation correspond to the \MC\ 
carrying the angular momentum with it and thereby altering the profile
of $v_{\phi}$. We can get an equation for specific angular momentum
(i.e.\ angular momentum per unit mass) ${\mathcal L} = r \sin \theta \:
v_{\phi}$ by multiplying (16) by $r \sin \theta$, which gives
$$\frac{\pa {\mathcal L}}{\pa t} + \vb_m. \nabla {\mathcal L}
= r\sin \theta \,\frac{K_{\phi}}{\rho}. \eqno(17)$$
Using the relation $v_{\phi} = r \sin \theta \: \Omega$,
it is also easy to cast (16) into the form of an equation of $\Omega$.
Readers will find that sometimes in the literature the equation of
azimuthal dynamics is written in terms of $\Omega$ rather than $v_{\phi}$. 
To solve (16), we need to evaluate
the turbulent stress $\rho \, \ol{v_i' v_j'}$ required for obtaining
$\Kb$ through (10).  Since the mathematical theory of the turbulent
stress is extremely complicated, we now discuss the basic physics
of the problem qualitatively without getting into the details of the
mathematical theory.

Within a stellar convection zone, hot blobs of gas move upward and
cold blobs of gas move downward.  We may naively expect these blobs
to carry their angular momentum with them when they move upward or
downward.  This suggests that angular momentum may get well mixed
within the convection zone, such that the specific angular momentum
is constant throughout the convection zone.  Taking $s = r \sin
\theta$ as the outward distance from the rotation axis, specific angular
momentum in a region of the convection zone would be ${\mathcal L} =
s^2 \Omega$. If this were to be constant throughout the convection
zone, then $\Omega$ would fall off as we go further from the rotation
axis. However, we find the opposite of this in the Sun, as seen
in Figure~3.  In order for
the equatorial regions of the Sun to have higher $\Omega$, 
 we need some mechanism to continuously pump angular
momentum away from the rotation axis so that it can get piled up in
the equatorial regions, making those regions to rotate faster. Let us
now consider what kind of turbulent stresses can do this. 

\begin{figure}
\centerline{\includegraphics[width=0.9\textwidth,clip=]{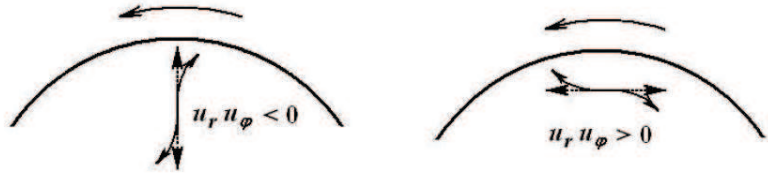}}
\caption{A sketch illustrating angular momentum transport 
in the equatorial plane by
turbulent mixing.  The direction of rotation is indicated at the
tops.  The left and right panels indicate how radially moving and
horizontally moving fluid blobs are deflected by the Coriolis 
force.  From Kitchatinov [40].}
\end{figure}

If the $\phi$-component of momentum $\rho v_{\phi}$ has to be 
advected in the radial direction, it is easy to argue that 
$\rho \, \ol{v_r v_{\phi}}$ would do the 
job and would contribute to a radial flux of angular momentum.
The crucial question is whether this will be positive or negative.
To address this question, it is convenient to look at the system
(i.e.\ the star)
from the frame of its average angular velocity. We had written
down our dynamical equation (11) with respect to an inertial frame.
When the variation of $\Omega$ over a star like the Sun is small
compared to the average value of $\Omega$, it is indeed often useful
to introduce a frame rotating with the average $\Omega$. It is well
known that, in such a frame, we shall have an additional Coriolis force
term $-\, 2 \, {\bf \Omega} \times \vb$ appearing on the right hand side of (11).
Now, look at the left panel of Figure~6,
indicating the direction of motion of a convective blob moving
radially (upward or downward) near the equatorial 
plane.  Assuming that we looking down from
the rotation axis, it is easy to show that the Coriolis force would
make the bob move as indicated in the figure.  Clearly  $\rho \, v_r v_{\phi}$
is negative for such a blob, indicating that radially moving convective
blobs would transport angular momentum downward. Presumably, such
transport would tend to make $s^2 \Omega$ constant within the
convection zone.  Now, consider a horizontally moving turbulent blob
shown in the right panel of Figure~6.  It is easy to check that
that the Coriolis force would make the blob move as shown in Figure~6,
leading to positive $\rho \, v_r v_{\phi}$.  We conclude that such
turbulent blobs would transport the angular momentum outward. 
Hence, in the solar convection zone, we shall have the required
outward pumping of angular momentum if horizontal turbulent motions
are more dominant within the solar convection zone compared to
radial turbulent motions.  Can this be the case under some
circumstances?  

While we do want to get into a full discussion of the complicated
problem of calculating turbulent stress tensors, we point out how the
deflections caused by the Coriolis force, as indicated in Figure~6,
enter into the calculation of $\ol{v_r v_{\phi}}$ for a weakly rotating star.    
Let us now consider a convective blob which would have the velocity
$$\vb_0 = v_{0,r} \, \eb_r + v_{0,\phi} \, \eb_{\phi}\eqno(18)$$
associated with it in the absence of the Coriolis
force. If the Coriolis force acts on the blob during its
coherence time $\tau$, then the velocity induced by the Coriolis
force will be
$$\vb_1 = - \, 2 \, {\bf \Omega} \times \vb_0 \, \tau \eqno(19)$$
so that the velocity with the Coriolis deflection becomes
$$\vb = \vb_0 + \vb_1. $$
Using (18) and keeping in mind that ${\bf \Omega} = \Omega \cos \theta \, \eb_r
- \Omega \sin \theta \, \eb_{\theta}$ at the colatitude $\theta$, from (19) we get
$$\vb_1 = 2 \, \Omega \, \tau \, v_{0,\phi} \sin \theta \, \eb_r + 2 \, \Omega \, \tau \,
v_{0, \phi} \cos \theta \, \eb_{\theta} - 2 \, \Omega \, \tau \, v_{0,r} \sin \theta \, \eb_{\phi}.
\eqno(20)$$ 
The turbulent stress term we are interested in is given by
$$ \ol{v_r v_{\phi}} = \ol{(v_{0,r} + v_{1,r})( v_{0,\phi} + v_{1,\phi})}
=\ol{v_{0,r} v_{0, \phi}} + \ol{v_{1,r} v_{0,\phi}} +
\ol{v_{0,r} v_{1,\phi}} + \ol{v_{1,r} v_{1, \phi}}. \eqno(21)$$
If we make the simplifying assumption in this discussion that 
$v_{0,r}$ and $v_{0,\phi}$ would be uncorrelated in the absence of the
Coriolis force, then $\ol{v_{0,r} v_{0, \phi}} = \ol{v_{0,r}} \; \ol{v_{0, \phi}}
=0$ Also, we expect $\Omega \, \tau$ to be small for weak rotation, so that
we can neglect $\ol{v_{1,r} v_{1, \phi}}$, which will be quadratic in 
$\Omega \, \tau$.  Substituting for
$v_{1,r}$ and $v_{1, \phi}$ from (20) into (21), we get
$$ \ol{v_r v_{\phi}} = 2 \, \Omega \, \tau \, (\ol{v_{0,\phi}^2} - \ol{v_{0,r}^2})
\sin \theta. \eqno(22)$$
It is clear that $\ol{v_r v_{\phi}}$ is positive when $\ol{v_{0,\phi}^2} > \ol{v_{0,r}^2}$, leading to outward transport of angular momentum, and is
negative when $\ol{v_{0,r}^2} > \ol{v_{0,\phi}^2}$, leading to inward transport of angular momentum, in conformity with the discussion accompanying
Figure~6.\footnote{This expression of $\ol{v_r v_{\phi}}$ was derived by 
Lebedinski [44] in his paper written in Russian.  I am grateful to Leonid
Kitchatinov for bringing this derivation to my attention.}

The crucial question now is whether $\ol{v_{0,\phi}^2} - \ol{v_{0,r}^2}$
appearing in (22) is positive or negative.  One standard result in
fluid mechanics is the Taylor--Proudman theorem (see [43], p.\ 183),
according to which fluid phenomena tend to be aligned parallel to
the rotation axis.  In accordance with this theorem, stellar
convection tends to take place in banana-shaped convection
rolls parallel to the rotation axis if the star rotates sufficiently
fast, as found in numerical simulations (see Figure~3 of [51] or
Figure~5 of [52]). If the star rotating slowly, we expect
radial turbulent motions to dominate.  This will lead to 
negative $\ol{v_r v_{\phi}}$ according to (22) and presumably a downward
pumping of angular momentum such that the equatorial regions may
rotate slower.  However, when a star rotates rapidly and banana-shaped
convection cells form, horizontal convective motions may become
more and more important making $\ol{v_r v_{\phi}}$ given by (22)
positive.  This is likely to cause outward pumping of angular
momentum, which may make the equatorial regions rotate faster.  Presumably,
something like this is happening in the Sun, although we should caution
the reader that the statement we just made is an over-simplification of
a complex situation. It is clear from (17) that the \MC\ also carries
angular momentum with it.  The final distribution of angular velocity
inside the convection zone follows from a complicated interplay of 
various angular momentum transfer terms.  Still, it is interesting
to note that numerical simulations
show that solar-like stars have anti-solar differential
rotation when rotating slowly and solar-like \DF\ when rotating
fast [51, 53], in qualitative agreement with the idea that, when stellar convection
is affected more by rotation, there is a higher tendency of angular momentum 
getting transferred outward. Whether rotation affects
stellar convection significantly depends on the dimensionless number $\Omega \, \tau$
appearing in (22). This dimensionless number (or
rather $2 \, \Omega \, \tau$) is often called the {\em Coriolis number} and
is essentially the inverse of what is known as the {\em Rossby number}.  A full
theory of turbulent stresses would involve calculating 
$\ol{v_{0,\phi}^2}/\ol{v_{0,r}^2}$ as a function of $\Omega \tau$,
which will make $\ol{v_r v_{\phi}}$ given by (22) a more complicated
nonlinear function of $\Omega \, \tau$.  We do not go into the details of
this complicated subject.  

Certainly (22) is not of the form (12).  In general, one should write
$$\ol{v_i' v_j'} = Q^{\Lambda}_{ij} - {\mathcal N}_{ijkl} \frac{\pa \ol{v_k}}
{\pa x_l} \eqno(23)$$
[49, 50]. The term $Q^{\Lambda}_{ij}$, which would incorporate expressions like (22),
is the essence of the $\Lambda$-effect, as indicated by the superscript $\Lambda$.
However, we would like to emphasize that, even by taking the
turbulent stress to be of the simple form (12),
Kippenhahn [47] succeeded in driving the \DF\ by 
assuming the coefficient $\mu_{T,r}$
for the radial viscous transport to be different from the coefficient
$\mu_{T,h}$ for the horizontal viscous transport.  We naively expect that
the viscous drag may oppose relative motions between fluid layers inside
a star, leading to solid body rotation.  This is indeed found to be the
case when $\mu_{T,r}=\mu_{T,h.}$  However, when these coefficients were
taken to be unequal, Kippenhahn [47] found the following intriguing results:  a
larger $\mu_{T,r}$ led to lower angular velocity near the equatorial
region, whereas a
larger $\mu_{T,h}$ led to higher angular velocity there. Presumably, the
case of radially moving fluid blobs discussed above corresponds to higher 
$\mu_{T,r}$, whereas the case of horizontally moving fluid blobs 
corresponds to higher $\mu_{T,h}$.  Thus, Kippenhahn's conclusions are
in agreement with the physics encapsulated in Figure~6 as we discussed
above. Afterwards, Kitchatinov and R\"udiger [49, 50] calculated the turbulent
stress tensor from their model of turbulence and constructed more detailed
models of stellar rotation.  They also found that angular velocity is 
higher in the equatorial regions (as we see for the Sun) when the turbulent
stress due to the horizontal motions dominates. We shall discuss these
solutions in \S3.4.

\subsection{Driving the \MC}

We are now ready to discuss on the basis of (15) how the \MC\ inside a
star is driven.  If $\Kb$ is taken to be as given in (13), then $\nabla
\times \Kb$ would equal $\mu_T \nabla^2 \omega$ and it is clear that
the last term in (15) would be a term giving dissipation of vorticity.
We easily see that (15) is an equation of the nature of (5), with the
first two terms on the right hand side of (15) as the source terms which
drive the \MC.  We now discuss the significance of these crucial source terms.

The term $- \nabla p \times \nabla \rho/\rho^2$ is called the {\em thermal
wind} term. Within the convection zone of a star like the Sun, the ascending
and descending convective blobs are deflected by the Coriolis force.  This
effect is less on the convective blobs moving near the polar regions. Due to this,
convective heat transport is expected to be more efficient in
the polar regions.  This is likely to make the polar temperature slightly
higher than the temperature at the equator.   It was realized in
the 1970s that the effect of rotation on convection may make the
heat transport latitude-dependent and that this may give rise to
large-scale flows [54, 55]. A
higher temperature (and a higher pressure) at the poles would drive a
\MC\ which is equatorward near the surface.  However, this is exactly
the opposite of what we observe! This means that the other source term
has to overcome this effect to drive the \MC\ in the correct direction.
Let us point out that the thermal wind term indeed mathematically leads
to a \MC\ opposite to what is seen the Sun.  The thermal wind term arises
when the contours of constant $\rho$ and constant $p$ do not coincide.
The solar surface can be taken as a surface of constant $\rho$.  If the
polar region is hotter, then a surface of constant $p$ which intersects
the solar surface at mid-latitudes would be above the solar surface near
the poles and would be below the solar surface near the equator.  It is
easy to check that $- \, \nabla p \times \nabla \rho$ will be positive in
the northern hemisphere.  It then follows from (15) that this term will
tend to drive a \MC\ with positive vorticity in the northern hemisphere,
opposite to what we find in the Sun, as pointed out in \S1.

\begin{figure}
\centerline{\includegraphics[width=0.6\textwidth,clip=]{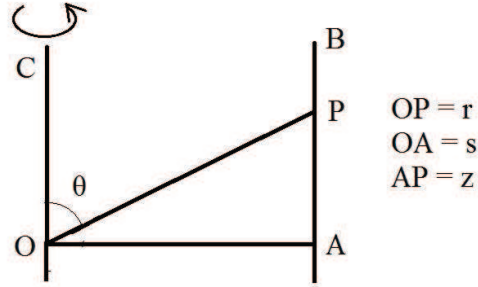}}
\caption{A sketch indicating the relation of the $(s, z)$ coordinates
with the $(r, \theta)$ coordinates.}
\end{figure}

Let us now turn our attention to the other source term 
$[\nabla \times \{ \vb \times (\nabla \times \vb) \} ]_{\phi}$ in (15).
This term is clearly quadratic in $\vb$. 
When we substitute the velocity field given by (2) and (3) in this,
we find that there are some terms which are quadratic in meridional
components $v_r$, $v_{\theta}$ and some terms which are quadratic in
$\Omega$.  We have already pointed out the velocities associated with
the \MC\ are much smaller than the velocities connected with the solar
rotation over much of the Sun.  We now make the approximation of keeping
only the terms quadratic in $\Omega$.  Then, a few lines of easy algebra
give us
$$[\nabla \times \{ \vb \times (\nabla \times \vb) \} ]_{\phi}
= r \, \sin \theta \, \cos \theta \, \frac{\pa}{\pa r} \Omega^2
- \sin^2 \theta \, \frac{\pa}{\pa \theta} \Omega^2 \eqno(24)$$
To understand the significance of this expression, let us consider
a straight line APB parallel to the rotation axis OC at a distance 
$s = r \, \sin \theta$ from it, as shown in Figure~7.  If $z$ is
measured upward from the equatorial plane OA, then $z = r \, \cos \theta$.
We can use $s$ and $z$ as our two independent spatial coordinates
in the place of $r$ and $\theta$.  This means that
$$\left( \frac{\pa}{\pa z} \right)_s 
= \left( \frac{\pa r}{\pa z} \right)_s \frac{\pa}{\pa r}
+ \left( \frac{\pa \theta}{\pa z} \right)_s \frac{\pa}{\pa \theta}
= \cos \theta \frac{\pa}{\pa r}
- \frac{\sin \theta}{r} \frac{\pa}{\pa \theta}. \eqno(25)$$
It then follows from (24) that
$$[\nabla \times \{ \vb \times (\nabla \times \vb) \} ]_{\phi}
= r \, \sin \theta \, \frac{\pa}{\pa z} \Omega^2.\eqno(26)$$
Substituting this in (15), we get
$$\frac{\pa \omega_{\phi}}{\pa t} =
 r \, \sin \theta \, \frac{\pa}{\pa z} \Omega^2 -
\frac{1}{\rho^2} \, [\nabla p \times \nabla \rho]_{\phi} 
+ \left[\nabla \times \left(\frac{\Kb}{\rho}
\right) \right]_{\phi}, \eqno(27)$$
which is our crucial equation.  We may point out that, if we had not
neglected the terms quadratic in the \MC\ velocities, then there would have been the
following additional term in the left hand side of (27)
$$+ s \nabla. \left( \vb_m \frac{\omega_{\phi}}{s} \right) \eqno(28)$$
which corresponds to the \MC\ carrying the vorticity with it.  

\begin{figure}
\centerline{\includegraphics[width=0.9\textwidth,clip=]{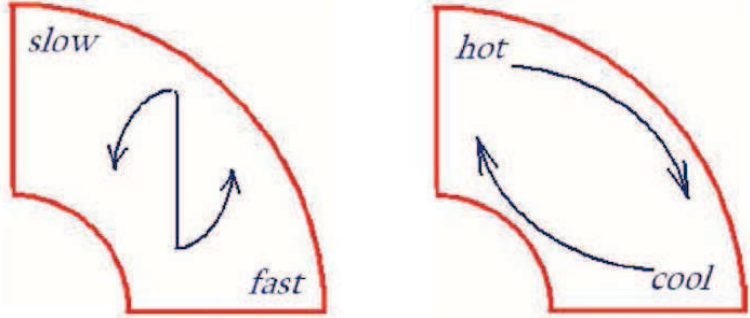}}
\caption{A figure indicating the directions in which the source terms
would tend to drive the \MC. The left panel indicates the centrifugal
term and the right panel the thermal wind term.  From Kitchatinov [41].}
\end{figure}

Let us now discuss the physical significance of the first source term
in the right hand side of (27). The appearance of $\Omega^2$ suggests
that this term may be connected with the centrifugal force, which turns
out to be the case. This term is naturally called the {\em centrifugal
term}. Suppose we consider a straight line parallel to 
the rotation axis inside the solar convection zone, like the line APB shown
in Figure~7.  It is obvious that the centrifugal force near the
equatorial region is larger than the centrifugal force at higher
latitudes, if the rotation profile inside the solar convection zone is
as shown in Figure~3---with higher $\Omega$ near the equatorial region.
If we subtract some mean centrifugal force averaged along the line APB,
then the net force near the equator would be in the outward direction
and the net force at higher latitudes in the inward direction.  This
would tend to drive a meridional circulation which is in the same
sense as the \MC\ of the Sun, as indicated in the left panel of Figure~8.
The right panel of Figure~8 shows the kind of \MC\ that the thermal
wind term would tend to drive.  It may be noted that, if $\Omega$ were
constant along lines parallel to the rotation axis like APB, then it
is easy write the centrifugal force as a gradient, showing that it is
a conservative force in this situation.  Only when $\Omega$ varies
with $z$, the centrifugal force becomes non-conservative and can
drive a circulation.  When we turn to mathematics after
understanding the physical concepts, it is easy to check that 
$\pa \, \Omega^2/ \pa z$ in the solar convection zone in the northern
hemisphere is negative, showing that the centrifugal term in (27) would tend to 
produce a \MC\ with negative vorticity, which is the case for the Sun
in the northern hemisphere.  

We thus conclude that the \MC\ in the Sun or similar stars arises
out of the interplay between the two source terms.  The thermal wind
term would try to drive a \MC\ in the sense opposite to what is seen
in the Sun.  Presumably, the centrifugal term overcomes this and drives
the \MC\ in the correct direction. Is it possible that the entire solar
surface is at the same temperature so that the thermal wind term is zero
and the centrifugal term alone drives the \MC\ of the Sun in the right
direction?  As shown in the Appendix, if 
we make an order of magnitude estimate for the solar
convection zone, the dissipation term (i.e.\ the last term) in (27)
turns out to be several orders of magnitude smaller compared to the
centrifugal term, when we use typical values of the large-scale flow
velocities in the Sun.  If the centrifugal term alone was driving
the \MC, then the centrifugal term arising out of the solar rotation
profile would drive a much stronger \MC.  The only possibility is that
the thermal wind term must be nearly comparable to the centrifugal
term and should balance it.  The small leftover part of the centrifugal
term must be driving the solar \MC. Kitchatinov and R\"udiger [50] estimated
that the solar pole has to be hotter by about 4 K compared to the equator
to give rise to a thermal wind term comparable to the centrifugal
term.  There have been some attempts to measure if there is any
temperature variation on the solar surface from the equator to the
pole [56, 57]. This is a difficult measurement and, although the results 
may not be completely conclusive, there are indications that the poles
of the Sun are indeed slightly hotter. We discuss in the Appendix how
an order of magnitude estimate of the pole-equator temperature difference
can be made.

It is an intriguing question why the two source terms in (27) are
comparable in magnitude.  Presumably, this is not an accident.  Let
us consider what would happen if the centrifugal term becomes much
larger.  Then it would drive a much stronger \MC. We see in (17) that
the \MC\ can carry angular momentum with it, changing the profile
of $\Omega$.  A stronger \MC\ would change the profile of $\Omega$
in such a manner that the centrifugal term given by (26) is reduced,
thereby decreasing the \MC.  We believe that there is such a feedback
mechanism in the Sun which keeps the two source terms in (27) comparable
in amplitude. 

We now show how to cast the thermal wind term in a different form 
involving the specific entropy $S$ per unit mass, since readers may
often encounter the thermal wind term written in this form
in the literature.  We have
$$\nabla p \times \nabla \rho = \left( \frac{\pa p}{\pa r} \eb_r + \frac{1}{r} 
\frac{\pa p}{\pa \theta} \eb_{\theta} \right) \times 
\left( \frac{\pa \rho}{\pa r} \eb_r + \frac{1}{r} 
\frac{\pa \rho}{\pa \theta} \eb_{\theta} \right) = \frac{1}{r}
\left( \frac{\pa p}{\pa r} \frac{\pa \rho}{\pa \theta} - 
\frac{\pa \rho}{\pa r} \frac{\pa p}{\pa \theta} \right) \eb_{\phi}.
\eqno(29)$$
For a parcel of gas, we have the basic thermodynamic relation
$$T \, dS = C_V \, dT + p \, d \left( \frac{1}{\rho} \right), \eqno(30)$$
where $C_V$ is the specific heat per unit mass. Eliminating $T$ by
using the ideal gas law $p = (\gamma -1) \, C_V \rho \, T$, (30)
can easily be put in the form
$$d \rho = \frac{\rho}{\gamma \, p} \, d p - \frac{\rho}{\gamma \, C_V} \, dS.
\eqno(31)$$
Substituting this for the differential of $\rho$ in (29), we arrive at
$$\nabla p \times \nabla \rho = \frac{1}{r} \frac{\rho}{\gamma \, C_V}
\left( \frac{\pa S}{\pa r} \frac{\pa p}{\pa \theta} - 
\frac{\pa p}{\pa r} \frac{\pa S}{\pa \theta} \right) \eb_{\phi}.
\eqno(32)$$
Now, convection tends to equalize entropy in the radial direction so that we
have
$$\frac{\pa S}{\pa r} \approx 0 \eqno(33)$$
within the convection zone.  Also, the hydrostatic equilibrium condition
is 
$$\frac{\pa p}{\pa r} = - \rho \, g, \eqno(34)$$
where $g$ is the acceleration due to gravity. Substituting (33) and (34) in
(32), the dominant term is
$$\frac{\nabla p \times \nabla \rho}{\rho^2} = 
\frac{1}{r} \frac{g}{\gamma \, C_V}\frac{\pa S}{\pa \theta} \, \eb_{\phi}.
\eqno(35)$$
Substituting this in (27), we get
$$\frac{\pa \omega_{\phi}}{\pa t} =
 r \sin \theta \, \frac{\pa}{\pa z} \Omega^2 -
\frac{1}{r} \frac{g}{\gamma \,  C_V}\frac{\pa S}{\pa \theta}
+ \left[\nabla \times \left(\frac{\Kb}{\rho}
\right) \right]_{\phi}. \eqno(36)$$
If the poles are hotter, then clearly $\pa S/\pa \theta$ is negative and
the thermal wind term tends to create positive vorticity in agreement
with our earlier discussion.

When the \MC\ is maintained in a steady by a balance between the two
large terms in (36), we have the thermal wind balance equation
$$r \sin \theta \frac{\pa}{\pa z} \Omega^2 =
\frac{1}{r} \frac{g}{\gamma \, C_V}\frac{\pa S}{\pa \theta}. \eqno(37)$$
Balbus et al. [58] pointed out that one can get a profile of the \DF\ matching
observations remarkably well by integrating (37) with the assumption
that the $S$ is constant over contours of constant $\Omega$ so that we
can write $S = f (\Omega^2)$.  This bypasses the need for evaluating
the turbulent stress terms.  However, the justifications for the assumption
$S = f (\Omega^2)$ do not appear particularly compelling to us.

\begin{figure}
\centerline{\includegraphics[width=0.9\textwidth,clip=]{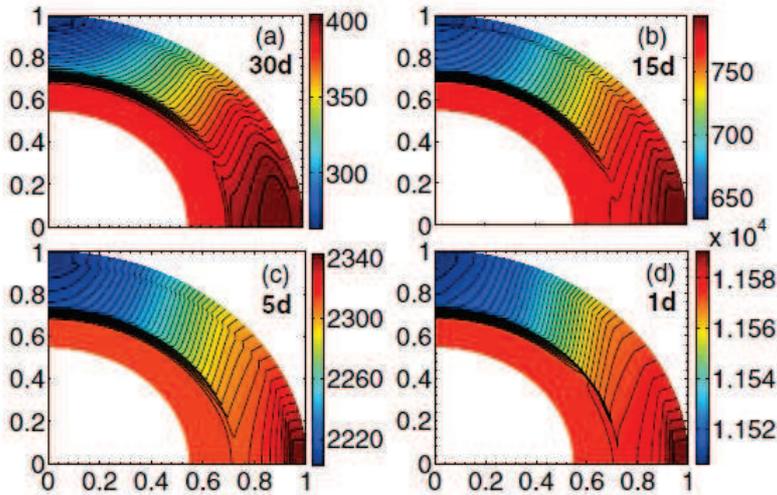}}
\caption{Theoretically computed profiles of angular velocity $\Omega (r, \theta)$  
in the poloidal planes of solar-mass stars with
rotation periods of 30, 15, 5, and 1 days. The rotational frequencies in nHz are
indicated by the different colours. From Karak, Kitchatinov,
and Choudhuri [62], based on the model of Kitchatinov and Olemskoy [61].}
\end{figure}

\subsection{Large-scale fluid flows inside solar-like stars}

In the previous subsections \S3.1--3, we have presented the basic physical
ideas of how we can theoretically calculate large-scale fluid flows
like the \DF\ and the \MC\ inside stars.  We basically need to solve (16) and (27)
simultaneously, with the time derivative terms set to zero when we deal
with a steady state, and accompanied by an equation for convective heat
transport to provide latitudinal variation of temperature that gives rise to the
thermal wind term. We often make the statement that the \MC\ is driven
by the turbulent stresses in the convection zone.  We should explain
what precisely we mean by this.  We have pointed out that turbulent stress terms
like $\ol{v_r v_{\phi}}$ estimated in (22) drive the \DF. While the turbulent
stresses may not explicitly appear in (27), they are the ultimate causes
of both the Coriolis term and the thermal wind term, the two drivers of the
\MC.  That is why we expect the \MC\ to be confined to the convection zone. 

The anisotropic viscosity model of Kippenhahn [47] gave rise to a
meridional circulation along with \DF\ due to the centrifugal term,
although the thermal wind term was not included in this model.  K\"ohler [59]
presented detailed computations of the \MC\ based on this model. As
we already pointed out, Kitchatinov and R\"udiger [50] calculated
both the \DF\ and the \MC\ based on their mean field model.
Due to many uncertainties in the parameters of the mean field theory,
it is difficult to say conclusively whether the \MC\ should consist of a single
cell in a hemisphere or should have a more complicated structure
[60]. For example, Kitchatinov and R\"udiger [50] found two
radially stacked cells of the \MC (see their Figure~1), whereas slight modifications in the model
led Kitchatinov and Olemskoy [61] to obtain a single cell. 
Now we briefly describe
some results presented by Karak et al. [62] based on the model of Kitchatinov
and Olemskoy [61].

As we pointed out in \S3.2, the nature of the \DF\ induced depends on
the nature of the turbulent stress.  If the star is weakly rotating,
then presumably radial turbulent motions dominate and $\Omega$ near the equator 
ends up with a lower value.  On the other hand, if the star is rotating fast,
horizontal turbulent motions may become more dominant and $\Omega$ near the equator 
ends up with a higher value. Figure~9 shows theoretically computed angular velocity 
patterns inside the convection zone of stars having mass equal to the 
solar mass, but rotating with different rotation periods [62].  All the
cases shown in Figure~9 correspond to situations in which the 
the rotation of the star is sufficiently fast and the equatorial
region has the higher $\Omega$ (like the Sun).  However, within this
regime, we see a clear trend.  If the rotation is made faster (i.e.\
rotation period shorter), then the contours of constant $\Omega$
tend to become cylinders parallel to the rotation axis.  On the other
hand, slower rotation tends to give contours constant over cones, as
in the Sun.  Presumably, the Sun is rotating fast enough to make the
horizontal turbulent motions important within the convection zone
so that $\ol{v_r v_{\phi}}$ given by (22) is positive, but
not fast enough to make $\Omega$ constant over cylinders. 

\begin{figure}
\centerline{\includegraphics[width=0.7\textwidth,clip=]{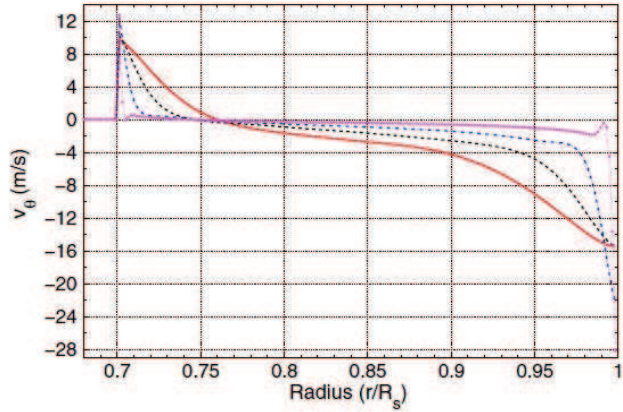}}
\caption{Theoretically computed component $v_{\theta}$ (in m s$^{-1}$) 
of meridional circulation at 45$^{\circ}$ latitude
of solar-mass stars with different rotation periods. 
Solid (red), dashed (black), dash-dotted (blue), and dot-pointed
(magenta) lines correspond to stars with rotation periods of 30, 15, 5,
and 1 days, respectively.  From Karak, Kitchatinov, and Choudhuri [62],
based on the model of Kitchatinov and Olemskoy [61].}
\end{figure} 

As $\Omega$ tends to become constant over cylinders for stars rotating
fast, it is obvious that $\pa \, \Omega^2/\pa z$ will tend to become
smaller.  Since the main driver of the \MC\ becomes weaker for faster
rotating stars, detailed computations show that the \MC\ is weaker
in faster rotating stars and tends to be confined to the edges of the
convection zone where the condition of constancy over cylinders is
expected to be violated in thin boundary layers. Some results [62] are
shown in Figure~10. As we shall discuss
later, this result that the \MC\ becomes weaker for faster rotating
stars poses some problems in modelling stellar dynamos. 

Although we do not intend to present a full discussion of numerical
simulations in this paper, we describe a few main results.  As already
pointed out in \S3.2, simulations of stellar convection showed that
slowly rotating stars have anti-solar \DF\ with the equatorial region
having lower $\Omega$ and rapidly rotating stars have solar-like \DF\
with the equatorial region having higher $\Omega$ [51, 53]. However,
rapidly rotating stars with accelerated equatorial regions tend to
have angular velocity constant over cylinders in most of the simulations,
indicating that the Taylor--Proudman constraint is quite strong.
Getting the angular velocity constant over cones (as found by helioseismology) rather than over cylinders has proved particularly
difficult in numerical simulations [63, 64].  If $\Omega$ is constant over
cylinders, then the centrifugal term given by (26) would be much smaller
than what it is inside the Sun and the \MC\ which one gets from such
simulations should be interpreted with caution.  Careful simulations
of the \MC\ showed that it is possible to get single cell \MC s for
slowly rotating stars with decelerated equatorial regions, but
rapidly rotating stars with accelerated equatorial regions tend to
produce multiple cells of the \MC\ [52, 53, 65].  A typical result is shown
in Figure~11. It may be noted that this result is based on an MHD code
including the dynamo action, leading to a variation of the \MC\ with
the solar cycle.  We shall discuss this problem in detail in \S5. If
we really have single-cell \MC\ in a hemisphere as indicated by the
most recent observational analysis [24], we have to confess that we
do not have simple, compelling arguments at the present time to explain
why it is so.

\begin{figure}
\centerline{\includegraphics[width=1.0\textwidth,clip=]{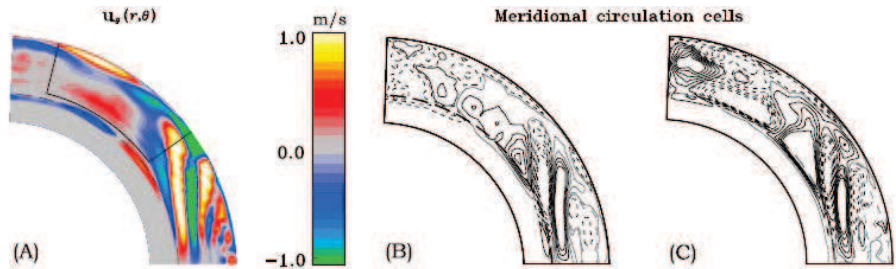}}
\caption{The mean \MC\ (averaged over turbulence) as found by Passos,
Charbonneau, and Miesch [65] from their numerical simulation of the solar convection.
(A) shows how the $\theta$ component of the \MC\ varies over the
meridional plane. This can be compared with the observational Figure~2.
(B) and (C) give streamlines of the \MC\ at the times of the solar
minimum and the solar maximum.  The dashed and solid lines respectively
indicate regions of negative and positive vorticity.}
\end{figure}

While discussing the basic mathematical theory of the \DF, we refrained
from a discussion of the boundary conditions
at the top and the bottom of the convection zone which we need to impose
while solving (16). If the observed conical isorotation contours have to
match with the solid body rotation in the radiation zone, then there has
to be a boundary layer at the interface.  The tachocline is such a boundary
layer.  Why the tachocline is so thin remains poorly understood.  Whether
the \MC\ or even magnetic fields play a role in keeping the tachocline
confined in a thin layer is an intriguing question [66, 67].
Rempel [68] developed a model of large-scale flows
by assuming simple forms of the turbulent
stress tensors and argued that the tachocline may play an important
role in breaking the Taylor--Proudman constraint even within
the convection zone.  A careful look at
Figure~3 shows a boundary layer of strong shear even at the solar
surface.  There have been attempts to model this shear layer though
simulations of the upper convection zone [64, 69].

\section{The role of \MC\ in solar dynamo models}

After discussing the relevant observations and basic theoretical ideas
connected with the \MC, we now turn our attention to the solar dynamo problem and
the role of the \MC\ in it. Because of the paucity of good pedagogical 
introductions to the theory of the \MC, we have discussed the basic theoretical
ideas about the \MC\ in a pedagogical manner in \S3.  There are, however,
convenient pedagogical introductions to dynamo theory ([43], Chapter 16; [70];
[71], Chapter 8; [72]) and also comprehensive reviews [73, 74, 75].  In view of this, 
our discussion of the basics of dynamo theory will be 
very brief, assuming that the readers are familiar with the fundamentals 
of MHD.  

\subsection{Basics of solar dynamo theory}

Just as an axisymmetric velocity field can be written in the form (1--3), an
axisymmetric magnetic field can be written as
$$\Bb = B_{\phi} (r, \theta, t) \, \eb_{\phi} + \nabla \times [A (r, \theta, t) \, \eb_{\phi}],
\eqno(38)$$
where $B_{\phi} (r, \theta)$ is called the {\em azimuthal magnetic field} and
$$\Bb_p = \nabla \times [A (r, \theta, t) \, \eb_{\phi}] \eqno(39)$$
is called the {\em poloidal magnetic field}.  The basic idea of dynamo theory is
that the toroidal and the poloidal fields sustain each other through a feedback
loop. As we shall discuss below, it is easy to see that the differential rotation
can stretch the poloidal field lines to create the toroidal field.  How the poloidal
field can be generated back from the toroidal field is more complicated.  A crucial
idea was due to Parker [76], who suggested that turbulent helical motions can twist
the toroidal field to produce the poloidal field.  Since the Coriolis force due
to the Sun's rotation would cause the convective blobs in the Sun's convection
zone to rotate, we clearly have helical turbulence there which could conceivably
twist the toroidal field to produce the poloidal field.

The basic evolution equation of the magnetic field in MHD is the well-known
induction equation
$$\frac{\pa \Bb}{\pa t} = \nabla \times (\vb \times \Bb) + \eta \, \nabla^2 \Bb, 
\eqno(40)$$
where
$$\eta = \frac{1}{\mu_0 \sigma}\eqno(41)$$
is often referred to as the {\em magnetic diffusivity}, $\sigma$ being the electrical
conductivity. To study the behaviour of the magnetic field inside a turbulent fluid,
we have to split both $\Bb$ and $\vb$ into a mean part and a fluctuating part as we
did for the velocity field in (7).  We write
$$\Bb = \ol{\Bb} + \Bb', \; \;  \vb = \ol{\vb} + \vb'. \eqno(42)$$ 
Substituting into (40) and averaging, we get
$$\frac{\pa \ol{\Bb}}{\pa t} = \nabla \times (\ol{\vb} \times \ol{\Bb}) 
+ \nabla \times {\mathcal E} + \eta \, \nabla^2 \ol{\Bb}, \eqno(43)$$
where
$${\mathcal E} = \ol{\vb' \times \Bb'} \eqno(44)$$
is known as the {\em mean EMF}. Just as the turbulent stress $\rho \, \ol{v'_i v'_j}$
appearing in (10) is crucial in the theory of large-scale flows, this mean EMF is crucial
in dynamo theory.  Steenbeck et al.\ [77] developed the systematic mean field theory
of MHD in a turbulent situation, from which ${\mathcal E}$ can be calculated.  If the turbulence
is isotropic, then ${\mathcal E}$ can be written in the form
$${\mathcal E} = \alpha \, \ol{\Bb} - \eta_T \, \nabla \times \ol{\Bb}, \eqno(45)$$
where
$$\alpha = -\, {1 \over 3} \ol{\vb'.(\nabla \times \vb')} \,
\tau, \; \; \; \;
\eta_T = {1 \over 3} 
\ol{\vb'.\vb'} \, \tau, \eqno(46)$$
where $\tau$ is the correlation time (see [43], \S16.5 for a derivation). 
It is obvious from the expression of $\alpha$
in (46) that $\alpha$ is a measure of the helical turbulence in the fluid.
Substituting (45) in (43), we arrive at
$$\frac{\pa \ol{\Bb}}{\pa t} = \nabla \times (\ol{\vb} \times \ol{\Bb}) 
+ \nabla \times (\alpha \, \ol{\Bb}) + (\eta + \eta_T) \, \nabla^2 \ol{\Bb}, \eqno(47)$$
Clearly $\eta_T$ is of the nature of a diffusion coefficient, and (46) makes it clear
that it arises out of turbulence.  This turbulent diffusion coefficient $\eta_T$ is
usually much larger than $\eta$, which can be neglected compared to $\eta_T$.  Also,
as we shall be dealing only with mean fields now onwards, we simplify the notation
by dropping the overline to indicate the mean, as we did from (11) onwards in \S3.
Then we write (47) as 
$$\frac{\pa \Bb}{\pa t} = \nabla \times (\vb \times \Bb) 
+ \nabla \times (\alpha \, \Bb) + \eta_T \, \nabla^2 \Bb. \eqno(48)$$
This is the fundamental equation of the turbulent dynamo.

We shall assume that both the mean velocity field and the mean magnetic field
are axisymmetric.  Then we can substitute (3) and (38) for $\vb$ and $\Bb$ in
(48).  A few steps of easy algebra give us the following evolution equations
of the toroidal and the poloidal fields
$$ \frac{\pa B_{\phi}}{\pa t} 
+ \frac{1}{r} \left[ \frac{\pa}{\pa r}
(r \, v_r B_{\phi}) + \frac{\pa}{\pa \theta}(v_{\theta} B_{\phi}) \right]
= \eta_T \left( \nabla^2 - \frac{1}{s^2} \right) B_{\phi} 
+ s \,(\Bb_p.\nabla)\Omega, \eqno(49)$$
$$
\frac{\pa A}{\pa t} + \frac{1}{s}(\vb_m.\nabla)(s A)
= \eta_T \left( \nabla^2 - \frac{1}{s^2} \right) A + \alpha \, B_{\phi},
\eqno(50)$$
where $s = r \sin \theta$.
We see in (49) that the source term for the toroidal field is $s(\Bb_p.\nabla)\Omega$
which corresponds to the stretching of the poloidal field by the \DF\ to
produce the toroidal field.   The source term for the poloidal field
in (50) is  $\alpha B_{\phi}$ which, in conjunction with the expression of
$\alpha$ given in (46), encapsulates Parker's idea of helical turbulence
twisting the toroidal field to produce the poloidal field [76].

When the first solar dynamo models were constructed in the 1960s and 1970s,
the existence of the \MC\ was not yet established and the dynamo modellers
of that era did not realize that such a flow may have important consequences
for the solar dynamo.  Also, the helioseismology results of $\Omega$ were not
available at time.  Dynamo modellers of that era would assume a `reasonable'
profile of $\Omega$ and then solve (49) and (50) after setting $\vb_m$ (and
its components $v_r$, $v_{\theta}$) to zero.  It was found that one can obtain
a dynamo wave propagating towards the equator if the condition
$$\alpha \frac{\pa \Omega}{\pa r} < 0 \eqno(51)$$
known as the {\em Parker--Yoshimura sign rule} is satisfied in the northern
hemisphere [76, 78].  It has been mentioned in \S2.1 that sunspots
appear at increasingly lower latitudes with the progress of the solar cycle,
leading to the butterfly diagram of sunspots shown by the shaded areas in Figure~1.
The equatorward propagation of the dynamo wave was believed to provide the 
theoretical explanation for this drift of sunspots with the solar cycle, and
the solar dynamo models of that era could give nice butterfly diagrams.
We are aware of only one paper of that era [79] which studied some effects
of the \MC\ on the solar dynamo and pointed out that the \MC\ could change
the appearance of the butterfly diagram.

While these older models of the solar dynamo seemed reasonably successful
at that time, certain new developments in solar physics made their inadequacies
clear, paving the way to the formulation of the flux transport dynamo model, in
which the \MC\ plays a crucial role.  We turn to these developments now. 

\subsection{The  flux transport dynamo model}

Large sunspots often appear on the solar surface in pairs, with the two 
members of the pair having opposite magnetic polarities [80].  The appearance of
such bipolar sunspot pairs is explained by Parker's famous idea of magnetic
buoyancy [81].  The line joining two sunspots in a pair is usually approximately
parallel to the solar equator, but with a tilt which tends to increase with
latitude in spite of a large statistical scatter [80, 82].  This dependence of the tilt
on latitude is called {\em Joy's law}. This tilt is produced by the action
of the Coriolis force on the rising magnetic flux tubes [83].

As seen in Figure~3, the \DF\ is concentrated in the tachocline at the bottom
of the solar convection zone.  We believe that this is the region where strong
toroidal field is produced due to the action of the \DF\ on the poloidal field.
Presumably, this toroidal field can be stored in a stable layer there and
parts of it break out in the form of magnetic flux tubes which rise through
the convection zone to produce sunspots at the solar surface.  Using the thin
flux tube equation [84, 85], simulations have been carried out to study the buoyant
rise of flux tubes through the convection zone to produce sunspots
[86, 87,  83, 88, 89]. These simulations fit observational
data well only if the magnetic field inside the flux tubes at the bottom of
the convection zone is taken to be of order $10^5$ G.  However, helical
turbulence will be unable to twist such strong magnetic fields and the
traditional $\alpha$-effect arising out of the $\alpha$-coefficient given
by (46) cannot be operational. The poloidal field has to be generated in
some other manner.

Recent solar dynamo models usually invoke an idea due to Babcock [90] and 
Leighton [91] for the generation of the poloidal field.  They pointed out
that, when tilted bipolar sunspots decay and the magnetic field in them
spreads around, the magnetic field from the sunspot at the higher latitude
contributes more in building up the overall magnetic field at higher
latitudes.  Like the traditional
$\alpha$-effect, the Babcock--Leighton process also can be described by an
$\alpha$-coefficient which is concentrated near the solar surface and
which appears in (50) in exactly the same manner.  
Surface observations of sunspot pair tilts suggest
that $\alpha$ due to the Babcock--Leighton process is positive in the
northern hemisphere.  When combined with profile of $\Omega$ determined
by helioseismology, it was found that the Parker--Yoshimura sign rule (51)
is not satisfied at the low latitudes where sunspots are seen.  This
suggests that the dynamo wave should propagate poleward, in contradiction
to the observations. We need something to turn around the dynamo wave.
Choudhuri et al.\ [92] showed that the \MC\ can do the job.

\begin{figure}
\centerline{\includegraphics[width=0.7\textwidth,clip=]{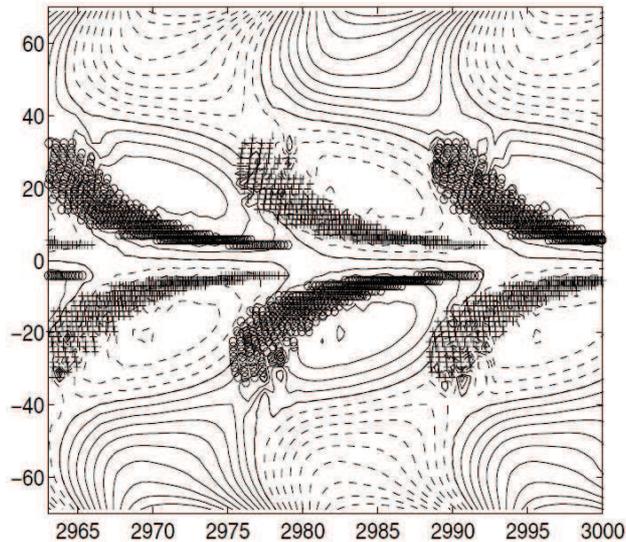}}
\caption{Theoretical butterfly diagram from a flux transport dynamo
simulation by Chatterjee, Nandy, and Choudhuri [103]. The latitudes where sunspots are
seen at a certain time are shaded, shown along with contours of constant
$B_r$ in the time-latitude plot.}
\end{figure} 

From the late 1980s, there were studies of how the poleward \MC\ near
the solar surface advects solar magnetic fields [6, 93, 94, 95], 
and Wang et al.\ [96] 
constructed a 1D dynamo model incorporating the \MC.  However, proper
2D models of the dynamo with the \MC\ were first constructed in 1995
by Choudhuri et al.\ [92] and Durney [97], and developed further in many subsequent
papers [98, 99, 100, 101, 102, 103].  If the toroidal
field is generated by the \DF\ in the tachocline and there is equatorward
\MC\ there, then the toroidal field can be advected equatorward, in spite
of the Parker--Yoshimura sign rule being violated, as shown
by Choudhuri et al. [92]. This would cause sunspots to form at increasingly 
lower latitudes with the progress of the solar cycle, whereas the poloidal
field near the solar surface is advected poleward by the poleward \MC\
there---in agreement with the observational data discussed in \S2.1. 
Figure~12 taken from Chatterjee et al.\ [103] presents a theoretical butterfly
diagram obtained from a dynamo model, along with contours of constant
$B_r$ at the solar surface in a time-latitude plot.  This theoretical 
figure can be compared favourably with the observational Figure~1.  Such
a remarkable agreement with observational data would be impossible
without incorporating the \MC\ in this kind of dynamo model, known as
the {\em flux transport dynamo model}. While different authors sometimes
use this term to mean slightly different things, we would refer to a
dynamo model as a flux transport dynamo model if the poloidal field
generation takes place by the Babcock--Leighton process and the \MC\
plays a crucial role in advecting the toroidal field at the bottom of
the convection zone and the poloidal field at the surface.  The \MC\
even decides the period of the dynamo cycle.  The dynamo period turns
out to be essentially equal to the time taken by a fluid element to
travel from higher latitudes to lower latitudes at the bottom of the
convection zone.  If the \MC\ is made stronger in the model, the period
becomes shorter [99].

\begin{figure}
\centerline{\includegraphics[width=0.6\textwidth,clip=]{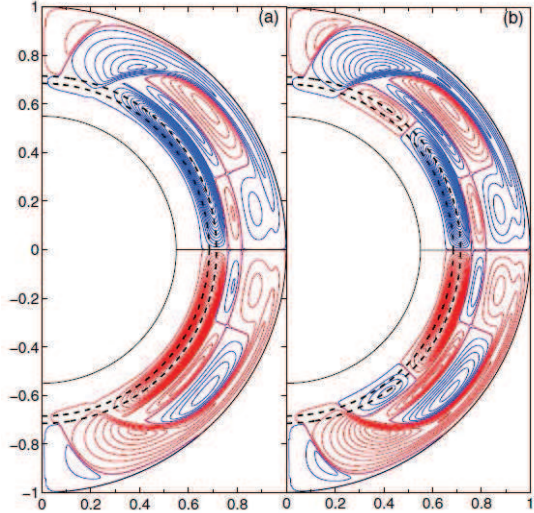}}
\caption{Two arbitrarily complicated \MC s with which Hazra, Karak, and Choudhuri [104]
solved the equations of the flux transport dynamo. Blue contours imply 
counter-clockwise circulation, whereas red contours
imply clockwise circulation. One obtains reasonable butterfly diagrams
with these \MC s.}
\end{figure} 

It may be pointed out that most of the dynamo models were worked out by
assuming a single-cell \MC\ encompassing one full hemisphere, with the
equatorward flow at the bottom of the convection zone.  We mentioned
in \S2.2 that the nature of the \MC\ deeper down inside the convection
zone remains uncertain.  Hazra et al. [104] addressed the important question
of whether the flux transport dynamo model can match observational data
if the \MC\ is more complicated. They solved the equations of the flux
transport dynamo for some arbitrarily complicated \MC s, two of which are
shown in Figure~13.  They concluded that the flux transport dynamo can
work as long as there is a layer of equatorward flow at low latitudes
at the bottom of the convection zone.  Jouve and Brun [105] also presented
solutions of the flux transport dynamo with multi-cell \MC.  If there is
sufficiently strong downward pumping as suggested in some convection
simulations, that also can give rise to an appropriate dynamo wave at
the bottom of the convection zone even if the return flow of the \MC\
occurs well above the bottom [106].  

The flux transport dynamo models work best if the \MC\ is assumed
to penetrate a little bit below the bottom of the convection zone where
the toroidal flux can be stored in a stable layer [102, 107]. How much penetration
is possible remains controversial---some authors arguing that the \MC\
cannot penetrate much into the stable layer [108], whereas others have argued
in favour of a considerable penetration [109]. It has
also been suggested that the \MC\ may play an important role in storing
the strong toroidal field in the stable layer underneath the bottom of
the convection zone [110].

With indications that many other stars have cycles like the Sun, one
important question is whether flux transport dynamos work in other
solar-like stars [111].  We have discussed in \S3 how the large-scale flow
patterns like the \DF\ and the \MC\ can be theoretically calculated
for stars rotating with different rotation periods.  Using such theoretically
computed flow patterns, Karak et al. [62] constructed flux transport dynamo
models of solar-mass stars rotating with different rotational velocities. 
They could explain such observational features as the enhanced activity
for faster rotating stars.  However, these models have some difficulty
in explaining the observational data that faster rotating stars have shorter
activity cycles.  As we pointed out in \S3.4, theoretical considerations
suggest that faster rotating stars have weaker \MC, which would lead to longer
cycle periods [112, 62].  Hazra et al.\ [113] have suggested that the  inclusion of the downward turbulent pumping may help in closing the
gap between observations and theory.

Although the 2D models of the flux transport dynamo have been reasonably
successful in modelling many aspects of the solar cycle, one limitation
of such models is that magnetic buoyancy and the Babcock--Leighton process
are inherently 3D processes.  They are treated in 2D models with rather
drastic approximations [100, 114, 115].  Of late, there have been attempts
of developing 3D kinematic models of the flux transport dynamo, in which
the large-scale flows are assumed to be given and the magnetic field is treated
in a 3D manner [116, 117, 118, 119]. Another approach of treating the non-axisymmetric
nature of the Babcock--Leighton process is to study the evolution of $B_r$ on
the solar surface, under the action of diffusion and the \MC.  
See Jiang et al.\ [120] for a survey of such surface flux transport models.
While these models can handle the Babcock--Leighton process at the solar
surface quite satisfactorily, they cannot treat the evolution of the
magnetic fields in the polar regions realistically by not including
the submergence of the \MC\ underneath the surface near the polar
regions [118]. There have also been efforts of combining 2D flux
transport dynamo model (in $r$ and $\theta$) with the 2D surface
flux transport model (in $\theta$ and $\phi$) [121].

\subsection{Modelling irregularities in the solar cycle}

We now turn to the important question of how the various irregularities
in the solar cycle arise (reviewed in [122]) and shall see that the \MC\ plays quite an
important role in this problem also.  We first point out how several
time scales in the flux transport dynamo are related, since an understanding
of this will be necessary for our discussions.

If $l$ is the thickness of the tachocline within which the magnetic diffusivity
is $\eta_{\rm tach}$, then the diffusion time within the tachocline is $l^2/
\eta_{\rm tach}$. In order for magnetic fields to be advected within the tachocline
by the meridional flow velocity of order $v$, the time scale $R_{\odot}/v$ of
the \MC\ has to be shorter than this. Since the magnetic diffusivity $\eta_T$
inside the convection zone is expected to be several orders of magnitude larger
than that in the tachocline, we expect the diffusion time scale $L^2/\eta_T$
within the convection zone 
($L$ is the thickness of the convection zone) to be much shorter than that within the tachocline.
We basically have two possible ordering of these various times scales
$$L^2/\eta_T < R_{\odot}/v < l^2/ \eta_{\rm tach}, \eqno(52)$$
or 
$$R_{\odot}/v < L^2/\eta_T <  l^2/ \eta_{\rm tach}, \eqno(53)$$
Dynamo models have been constructed both satisfying (52) [103, 123, 124] 
and satisfying (53) [99, 125].
As long as we are interested only in modelling periodic features of the solar
cycle, both types of models had reasonable success.  However, when we introduce
fluctuations in the dynamo model for modelling the irregularities of the cycle,
models satisfying (52) and (53) behave very differently [124, 126], as we shall discuss
below. 

Since magnetic stresses can quench the flows driving the dynamo (treated in kinematic
models by introducing some heuristic quenching terms), the dynamo problem is essentially
nonlinear.  It was initially thought that the nonlinearities cause the irregularities
in the dynamo cycles [127].  After it was realized that the most obvious types of
nonlinearities cannot produce sustained irregularities, the attention in the last few
years has been turned to stochastic fluctuations in the dynamo [128].  However, 
the nonlinearities are probably responsible for certain kinds of 
irregularities. For example, the Gnevyshev--Ohl rule that the odd cycle tends to
be stronger than the preceding even cycle is likely be a manifestation of period
doubling in a nonlinear system [129, 130]. 

Let us now turn to the possible sources of stochastic fluctuations in the solar
dynamo.  One finds a scatter in the tilt angles of sunspot pairs around the mean
tilt satisfying Joy's law [82]. While the mean tilt results from the action of the
Coriolis force on rising flux tubes [83], the scatter is presumably caused by the buffeting
due to turbulence when the flux tubes rise through the convection zone [131]. This
implies that the Babcock--Leighton process for generating the poloidal field 
from tilted bipolar sunspots
should involve fluctuations [123].  If such fluctuations are introduced in theoretical
dynamo models satisfying (52), then their effects spread through the convection
zone in a few years.  On the other hand, fluctuations introduced in models satisfying
(53) do not diffuse much, but get carried with the \MC.  The strength of the cycle~24
predicted by Dikpati and Gilman [125] based on a dynamo model satisfying (53)
completely failed to match observations, whereas
the strength predicted by Choudhuri et al. [123] based on a dynamo model satisfying
(52) turned out to be the first successful dynamo-based prediction of a solar cycle
before its onset. The higher turbulent diffusivity of the convection zone corresponding to the
condition (52) also helps in explaining the preferred dipolar parity of the Sun [103, 132]
and the lack of hemispheric asymmetry [133, 134].  It appears that the solar situation
corresponds to the condition (52) rather than (53).  Dynamo models with stochastic
fluctuations in the Babcock--Leighton process can also produce
grand minima like the Maunder minimum [135].

As we pointed out in \S2.4, there is evidence of random fluctuations in the \MC\
of the Sun having correlation time of order 30--40 yr.  This is a second important 
source of fluctuations which is expected to affect the dynamo.  Let us try to figure
out as to what will happen if the \MC\ becomes weaker during an epoch due to these
fluctuations.  As discussed in \S4.2, a slower \MC\ will make the period of the dynamo
longer.  This will give rise to two competing effects.  The diffusion will have a longer
time to act, thereby trying to make the magnetic fields weaker.  The \DF\ also will have
a longer time to produce a stronger toroidal magnetic field.  Which of these two effects
dominates will depend on whether the condition (52) or the condition (53) is satisfied.
If the condition (52) holds, then diffusion will be dominant and longer cycles will be
weaker.  On the other hand, if the condition (53) holds, then the \DF\ generating more
toroidal field will be the dominant process, making longer cycles stronger.  Observational
data indicate that it the first possibility---longer cycles are weaker---which holds
for the Sun, again suggesting the condition (52) is the appropriate condition for the
Sun.  One observational fact known as the {\em Waldmeier effect}---that shorter cycles rise
faster---is a consequence of this.  If shorter cycles are stronger, they are certainly
expected to rise faster.  Only by considering fluctuations in the \MC\ causing the
durations of different cycles unequal, it has been possible to provide a theoretical
explanation of the Waldmeier effect [36].

Taking the fluctuations in the \MC\ to be the only fluctuations in the solar dynamo
process, Karak [136] succeeded in modelling the irregularities of the solar cycle to
some extent. Since fluctuations in the Babcock--Leighton process are also present, a
full theory should be based on the combined effect of both of these types of fluctuations.
Choudhuri et al. [123] made their prediction for cycle 24 at a time when the importance
of fluctuations in the \MC\ was not realized and these fluctuations were not taken into
account.  Presumably, this prediction turned out to be so successful because there was
no big random change in the \MC\ between the time of the prediction and the peak of the
next cycle.  A prediction of a future cycle should take into account of both fluctuations
in the Babcock--Leighton process and fluctuations in the \MC.  Observational data
suggest that changes in the \MC\ may take a few years to change the strength of the solar
cycle [137].  This delay in the effect of the \MC\ enables us to use the value of
the \MC\ at the end of a cycle (from the rate of decline of the cycle at that time) 
which is appropriate for determining the strength of the next cycle.  Also, the poloidal
field at the end of the cycle provides information about the fluctuations in the
Babcock--Leighton process that is needed for predicting the next cycle.  Hazra and
Choudhuri [138] have developed a formula for predicting the next cycle by using the
decline rate and the poloidal field at the end of the previous cycle.

At last, we come to the question of explaining the most extreme events in the irregularities
of the solar cycle---the grand minima when sunspots may disappear for several decades.
From the analysis $^{10}$Be concentration in polar ice cores, it has been possible to infer
that there had been about 27 grand minima in the last 11,000 years [139]. If the
fluctuations in the Babcock--Leighton process make the poloidal field at the end of
a cycle too weak, or if the fluctuations in the \MC\ make it too slow (keep in mind that
a slower \MC\ leads to a longer cycle of weaker strength), then that may drive the
dynamo into a grand minimum.  Dynamo simulations suggest that it is possible to produce
grand minima by considering fluctuations in the Babcock--Leighton process alone [135]
or in the \MC\ alone [136], if the fluctuations are assumed to be sufficiently large.
Considering both kinds of fluctuations simultaneously and choosing 
the statistical parameters of these
fluctuations on the basis of past observations, Choudhuri and Karak [140] succeeded
in explaining the statistical properties of grand minima reasonably well. Presumably,
the grand minima are produced by the combined effect of fluctuations in both the 
Babcock--Leighton process and the \MC\ [140, 141]. 
 
\section{Back reaction of the dynamo on the \MC}
 
We presented our discussion of large-scale fluid motions in \S3 by 
assuming that there is no
magnetic field present. If a magnetic field is present 
in the fluid, then it gives rise to the Lorentz force,
which has to be included in the basic dynamical equation (11) such that it becomes
$$\frac{\pa \vb}{\pa t} + \nabla \left( \frac{1}{2} v^2 \right)
- \vb \times (\nabla \times \vb) = - \,
\frac{\nabla p}{\rho} +  \Fb + \frac{\Kb}{\rho}
+ \frac{(\nabla \times \Bb) \times \Bb}{\mu_0 \, \rho}. \eqno(54)$$
The Lorentz force term can be written as
$$\Fb_L = \frac{(\nabla \times \Bb) \times \Bb}{\mu_0 \rho}
= - \, \frac{1}{\rho} \nabla \left( \frac{B^2}{2 \mu_0} \right)
+ \frac{(\Bb. \nabla) \Bb}{\mu_0 \, \rho}. \eqno(55).$$
The first term on the right hand side indicates that the magnetic field has a pressure
$B^2/ 2 \mu_0$ associated with it.  The other term is of the nature of {\em magnetic tension}.
If magnetic field lines are straight and parallel in a region, it is easy to see that
$(\Bb. \nabla) \Bb$ will be zero in that region.  This term arises when the magnetic field
lines are bent and tries to straighten the field lines. Another effect of the magnetic
tension is that it tries to shorten the lengths of magnetic field lines.

\begin{figure}
\centerline{\includegraphics[width=0.8\textwidth,clip=]{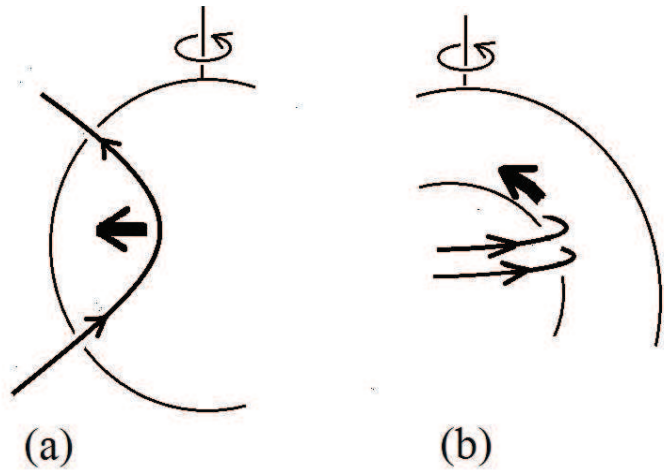}}
\caption{Sketches of (a) a typical magnetic field line inside the Sun and (b) a
band of strong toroidal field at the bottom of the solar convection zone. The
thick arrows indicate the parts of the Lorentz force which drive (a) torsional
oscillations and (b) variations in the \MC\ with solar cycle.}
\end{figure} 

The magnetic field generated in the Sun by the dynamo action will certainly have a
Lorentz force associated with it. 
This Lorentz force, appearing in (54) as shown above, would affect both the turbulent motions and
the mean motions.  The analytical theory of how turbulent motions are affected by the
Lorentz force is extremely complicated [142, 143].  Here we focus our attention on
the action of the Lorentz force on the large-scale mean motions.  In \S2.4 we mentioned
torsional oscillations and variations of the \MC\ with the solar cycle.  Before getting
into the mathematical theory of these, let us present the main physical ideas qualitatively. 
One important ingredient of the dynamo process is that the poloidal field lines get
stretched by the differential rotation to create the toroidal field.  We expect a
field line to look as shown in Figure~14(a). Such a field line would have a tension
force in the direction of the thick arrow, inducing motions in the azimuthal direction.
We believe that this is how torsional oscillations are driven.  To figure out the effect
of the magnetic field on the \MC, we keep in mind that the toroidal magnetic field in
the tachocline at
the bottom of the convection zone, as shown in Figure~14(b), would be the dominant
component of the magnetic field at the time of the solar maximum.  Due to magnetic
tension, the toroidal field lines will try to shorten their lengths.  Motions in the
radial direction would be inhibited if the layer of tachocline where the toroidal field
is stored is stable.  The easiest way for the toroidal field lines to shorten their lengths
is to slip in the poleward direction, which means that we would have a force
as indicated by the thick arrow in Figure~14(b).  This force would clearly oppose
the equatorward \MC\ at the bottom of the convection zone, causing a decrease in the
\MC\ speed at the time of the solar maximum, as seen in Figure~4. We now turn to
a discussion of the mathematical formulation of these qualitative ideas. 

To calculate the back reaction of the dynamo-generated magnetic fields on the large-scale
flows crucial for driving the dynamo, we need to solve (54) along with the dynamo
equations (49) and (50), which give the magnetic fields required to evaluate
the Lorentz force term in (54). Rempel [144] followed such a procedure and showed that the
back reaction gives rise to torsional oscillations and variations in the \MC.  Here
we shall discuss the action of the Lorentz force on azimuthal motions and meridional
motions separately, since that makes the basic physics of the problem clearer. One simple
way of incorporating the fact that the \MC\ becomes weaker when there are strong magnetic
fields is to include a quenching due to magnetic fields in the expression of the \MC\ [145].
Such a quenching has a tendency of making dynamos satisfying condition (53) unstable,
adding additional support to our contention that the condition (52) is the appropriate
condition for the solar dynamo.

Let us first consider the azimuthal motions which may be driven by the Lorentz force.  
We have to consider the $\phi$ component of (54).  This is nothing but (16) with an
additional term corresponding to the $\phi$ component of $\Fb_L$ given by (55). If we
calculate $(\nabla \times \Bb) \times \Bb$ by using the expression (38) of the magnetic
field, this additional term is found to have an elegant form
$$(\Fb_L)_{\phi} = \frac{1}{\mu_0 \, \rho \, s^3} \, J \left( \frac{s B_{\phi}, sA}{r, \theta}
\right) \eqno(56)$$
involving a Jacobian which essentially is a product of terms having toroidal and
poloidal components. This is in agreement with Figure~14(a) which suggests that the Lorentz
force driving azimuthal motions should involve both the toroidal and the poloidal components.
Chakraborty et al. [34] solved the dynamo equations (49) and (50) along with (16) with
the additional term given by (56).  They were able to develop a model of torsional oscillations
which agreed reasonably well with observational data.  Earlier theoretical efforts are
cited in this paper. 

We now turn to the problem which is of central interest to us: how variations in the
\MC\ with the solar cycle are produced by the Lorentz force. For this purpose, we have
to take the curl of (54) and consider its $\phi$ component. This leads to (27) with
the additional term $(\nabla \times \Fb_L)_{\phi}$. Taking the magnetic field as given
by (38) and using the expression (55) for $\Fb_L$, we find that the dominant terms in
the expression of this additional term are
$$(\nabla \times \Fb_L)_{\phi} = \frac{1}{\mu_0 \, \rho} \left[ \frac{1}{r^2} \frac{\pa}
{\pa \theta} (B_{\phi}^2) - \frac{\cot \theta}{r} \frac{\pa}{\pa r} 
(B_{\phi}^2) \right]. \eqno(57)$$
We note that the dominant terms in the part of the Lorentz force that causes variations
in the \MC\ arise from the toroidal component, in accordance with Figure~14(b). To find
out how the \MC\ varies with the solar cycle, we need to solve the dynamo equations
(49) and (50) along with (27) with the additional term given by (57). Hazra and 
Choudhuri [146] solved this problem by following a perturbative approach, in which
$\vb_m$ and its vorticity $\omega_{\phi}$ were divided into a time-independent average part
denoted by subscript 0 and a part varying with the solar cycle denoted by subscript 1,
i.e.
$$\vb_m = \vb_0 + \vb_1, \; \; \; \omega_{\phi} = \omega_0 + \omega_1. \eqno(58)$$
Substituting this in (27) with the additional term given (57) and subtracting from
it the equation for $\omega_0$, we end up with the equation for the perturbed part
$$\frac{\pa\omega_1}{\pa t} \, + \, s \, \nabla. \left( \vb_0 \frac{\omega_1}{s} \right)
+ \, s \, \nabla. \left( \vb_1 \frac{\omega_0}{s} \right) = 
\frac{1}{\mu_0 \, \rho} \left[ \frac{1}{r^2} \frac{\pa}
{\pa \theta} (B_{\phi}^2) - \frac{\cot \theta}{r} \frac{\pa}{\pa r} 
(B_{\phi}^2) \right] + \left[\nabla \times \left( \frac{\Kb_1}{\rho} \right) 
\right]_{\phi}. \eqno(59)$$
Note that we have included the small term given by (28), since we are dealing
with the equation of a small perturbed quantity.  We have also assumed that the
turbulent stress term $\Kb$ can be written in a form linear in the mean velocity as
in (13) and write the part of $\Kb$ associated with $\vb_1$ as $\Kb_1$. It is
clear from (59) that a part of the Lorentz force associated with the toroidal
magnetic field causes the variations in the \MC\ with the solar cycle.  Solving
the full equation (27) with the additional term (57) would be a particularly challenging
problem, since it would involve evaluating the thermal wind term which requires
thermodynamics in addition to fluid mechanics.  When we subtract the equation for
$\omega_0$ from the full equation, the thermal wind term drops out and it becomes
a much more tractable problem.  Hazra and Choudhuri [146] solved (59) with the dynamo
equations (49) and (50) to develop a theory of the variations of the \MC\ with the
solar cycle. 

\begin{figure}
\centerline{\includegraphics[width=1.0\textwidth,clip=]{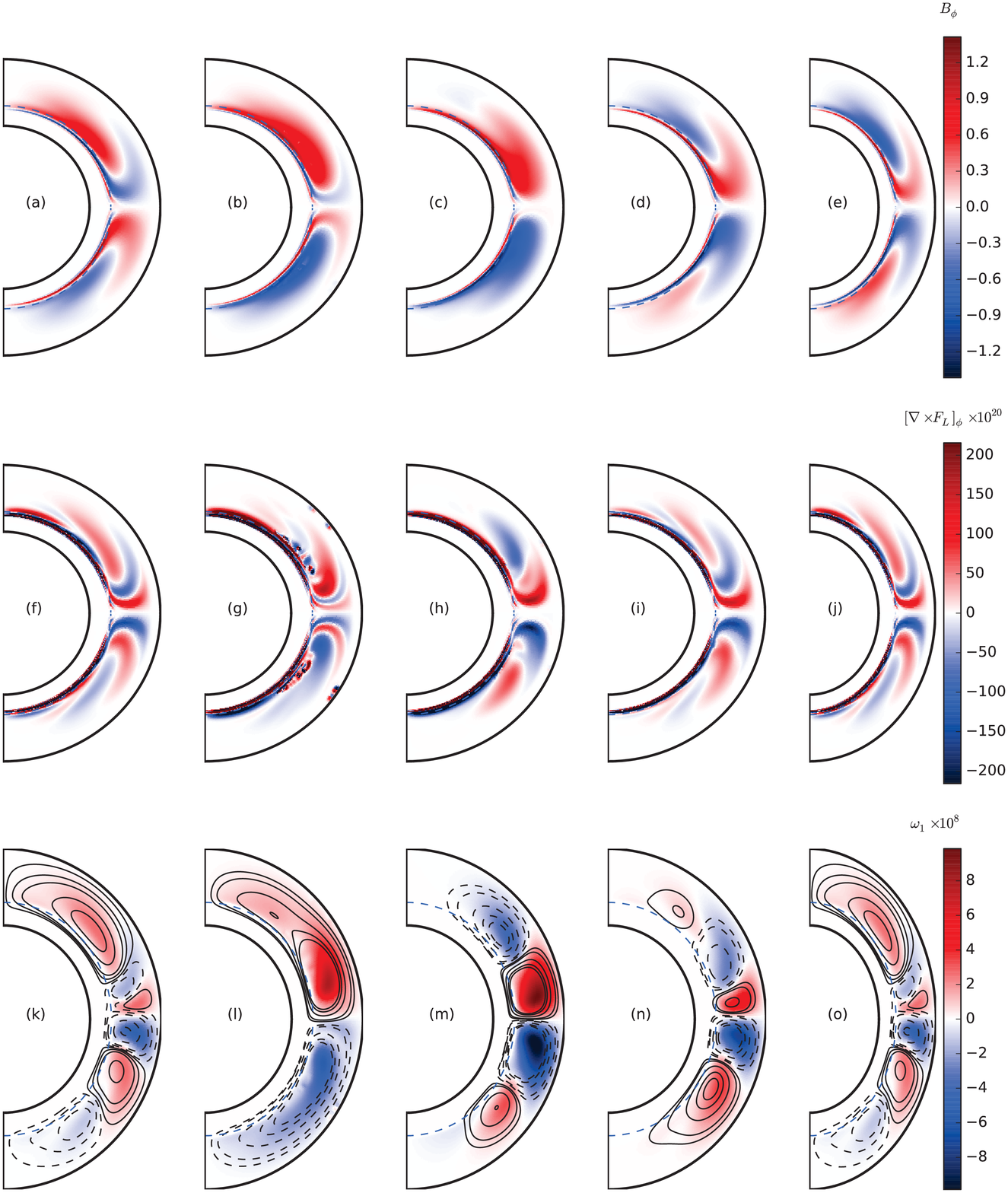}}
\caption{The evolution within the solar convection zone of the toroidal
magnetic component $B_{\phi}$ (top row), the part 
$(\nabla \times \Fb_L)_{\phi}$ of the Lorentz force driving the
variations in the \MC\ (middle row) and the perturbed vorticity $\omega_1$ associated with
the \MC\ variations (bottom row). The successive columns correspond to the
profiles at intervals of $T/8$ ($T$ is the dynamo period), the
second column corresponding to the solar maximum
and the fourth column to the solar minimum.  From Hazra and Choudhuri [146].}
\end{figure} 

Let us now point out one puzzle, which still remains unresolved. It is clear 
from (56) that the part of the Lorentz force driving torsional oscillations is
quadratic in toroidal and poloidal components, whereas (57) indicates that the
part driving variations in the \MC\ involves a simple square of the toroidal
component.  Since the toroidal component is much stronger than the poloidal
component in the Sun, we conclude that the driver of the variations in the \MC\ is
much stronger than the driver of the torsional oscillations.  We then expect the
variations in the \MC\ to have a much larger amplitude than that of the
torsional oscillations.  Observationally, however, both these amplitudes are
found to be comparable---of order 5 m s$^{-1}$ in the top layers of the convection
zone. Even simple order of magnitude estimates suggest that the variations in
the \MC\ with the solar cycle should be much larger than what they are [146].
We still do not have a resolution of this puzzle.  If the Lorentz
force appearing in (59) is divided by a suitable factor, then our theoretical
model gives variations in the \MC\ agreeing with observational data reasonably
well.

Since $\omega_0$ for the \MC\ in the northern hemisphere is negative, we want
$\omega_1$ to be positive at least in some regions of the northern hemisphere
at the time of the solar maximum so that the \MC\ becomes weaker at that
time.  Figure~15 taken from Hazra and Choudhuri [146] shows how $B_{\phi}$,
$(\nabla \times \Fb_L)_{\phi}$ and $\omega_1$ vary during the solar cycle, the
second column corresponding the solar maximum. It can be seen that the relevant
part $(\nabla \times \Fb_L)_{\phi}$ of the Lorentz force and $\omega_1$ driven by
it are both predominantly positive in the northern hemisphere at that time.  We
expect this to weaken the \MC\ at the time of the solar maximum.  Figure~16 shows
how the \MC\ at the mid-latitude at the surface varies with time, along with
the sunspot number calculated from the theoretical dynamo model.  This figure
compares favourably with Figure~4.

\begin{figure}
\centerline{\includegraphics[width=0.6\textwidth,clip=]{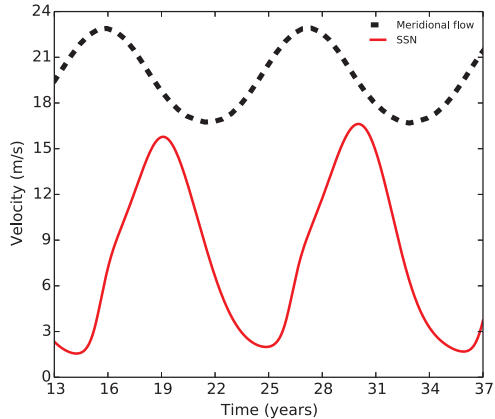}}
\caption{The variation of the \MC\ just below the solar surface at latitude
$25^{\circ}$ (black dashed line), as computed from a theoretical model.
The theoretical sunspot number (red solid line) is also shown. From Hazra
and Choudhuri [146].}
\end{figure} 

MHD simulations of stellar convection zone dynamics give rise to dynamo
cycles and the \MC\ varying with these cycles. This is clearly seen in Figure~11.
The variation of the \MC\ with the dynamo cycle was studied very carefully and
thoroughly by Passos et al. [147]. Apart from directly exerting a Lorentz force
on the large-scale flows like the \DF\ and the \MC, the dynamo-generated magnetic
fields can also modify the turbulent stresses which drive the large-scale flows.
This second effect, which is usually not included in the mean field models, is
automatically taken into account in the MHD simulations.  However, these
simulations so far have certain other limitations.  These MHD simulations
have demonstrated that dynamo action indeed takes place in stellar convection
zones and it is possible to get periodic solutions.  However, so far these
simulations have not yielded solutions which can be regarded as realistic
representations of the flux transport dynamo.   These simulations are still
far from matching actual solar cycle data, which a mean field
model can do with a suitable specification of parameters,
as seen in Figure~12.  The \MC\ obtained
in these simulations also has a multi-cell structure as shown in Figure~11.
Simulations can study the variations in such a \MC\ with the dynamo cycle,
even though either the dynamo cycle or the mean \MC\ may not look
very solar-like.  The advantage of the mean field
approach outlined in this Section is that one can choose the various parameters
appropriately to have solar-like cycles and solar-like $\vb_0$, from which
the time-varying part $\vb_1$ can be calculated.  Thus, both the mean field
approach and the simulations approach have their relative advantages and
limitations while modelling the variations of the large-scale flows in the
Sun.  Both these complementary approaches should be pursued to gain a deeper
insight into this complex problem.

There have been efforts of explaining the variations of the \MC\ with
the solar cycle on the basis of inward flows towards active region belts---presumably
driven by the fall in gas pressure in such belts [35]. Now that such cycle
variations of the \MC\ have been confirmed observationally even at the bottom
of the convection zone [24], such an explanation based on a local surface phenomenon
does not appear convincing to us.

\section{Conclusion}

This review focuses on the \MC\ of the Sun---driven presumably by turbulent
stresses in the solar convection zone.  Well before the current era of 
research in this field, Eddington [148] and Sweet [149] pointed out the possibility
of \MC s inside rotating stars.  Due to the polar flattening in a rotating
star, the temperature gradient tends to be steeper in the polar region. It may
not be possible to reconcile this with the nuclear energy generation process
without incorporating a \MC\ in the star, even in the radiatively stable regions.  Since
the rotational flattening of the Sun is very small, the Eddington--Sweet
circulation inside the Sun would be very slow with a time scale of order
$10^{12}$ yr---much larger than the age of the Universe ([150], \S42.5). The
\MC\ that we observe in the Sun is certainly a very different thing.

The \MC, which was first observed at the solar surface, is expected to be
confined within the convection zone.  This circulation is now
realized to be an important component of the solar dynamo process which generates
the solar magnetic field and its cycle. We believe that such a circulation exists in 
other solar-like stars as well, in which the dynamo cycles are probably generated
in the same manner [111].  Even for compact stars like neutron stars accreting matter
from a companion, flows in the meridional plane play a crucial role in the
evolution of their magnetic fields [151, 152].  While helioseismology has
thrown considerable light on the nature of the \MC\ underneath the solar surface,
its nature in lower regions of the convection zone still remains uncertain,
although recent results support the view that the equatorward flow exists at the bottom
of the convection zone [24].
Theoretical dynamo models work best if the \MC\ is assumed to have a single
cell spanning the whole of the convection zone in a hemisphere, although
more complicated circulations satisfying certain criteria can also be
accommodated [104]. 

The theoretical discussions in this review are primarily based on 2D mean
field models, since such models make the physics of the problem clear.
The theory of the \MC\ is intimately connected with the theory of the other
large-scale fluid flow pattern inside the Sun: the differential rotation. The
Coriolis force due to the Sun's rotation induces horizontal motions within the
convection cells, which may give rise to a transport of angular momentum away
from the rotation axis---leading presumably to a faster rotating equatorial region. Such
a pattern of differential rotation gives rise to a centrifugal term driving
the \MC\ in the direction consistent with observations.  However, this term
is opposed by a thermal wind term arising out of the fact that the Sun's
poles are probably slightly hotter (about 4 K according to [50]) due to the more efficient
convection in the polar regions. It seems that these two terms are comparable
in magnitude and a slight imbalance between them drives the \MC. 

Solar dynamo models initially started being developed at a time when even the
existence of the \MC\ was not known.  The early models without the \MC\ had
various difficulties which led to the formulation in the 1990s of the flux
transport dynamo model, in which the \MC\ plays a central role and even
determines the period of the dynamo cycle.  Irregular fluctuations in the
\MC\ (which seem to have a coherence time of about 30--40 yr [36]) are
important in explaining many aspects of the irregularities in the solar cycle,
in making comprehensive models of grand minima [140] and in predicting future
cycles [138 ]. The Lorentz force of the magnetic fields generated by the 
dynamo can react back on the large-scale flows like the \DF\ and the \MC\
causing their periodic variations with the solar cycle.

Since the \MC, the \DF\ and the dynamo action are all related to each other,
a full 2D mean field model should require simultaneous solution of (16), (27),
(49) and (50) with the additional terms (56) and (57) added to (16) and (27)
respectively. Since a calculation of the thermal wind term in (27) needs a realistic
model of convective heat transport in which the effect of the Coriolis force
on convection cells is included, an equation of heat transport also has to be
solved along with the equations listed above.  This is a formidable problem.
Much of our theoretical understanding of this field has come from solutions
of parts of this full problem.  This review focuses on such studies of parts
of the full problem, which elucidate many aspects of basic physics. In spite
of the major advances in the last few years, many issues remain poorly understood.
We hope that in the near future observations, mean field models and numerical
simulations will go hand in hand to solve many of the remaining puzzles.

\bigskip
{\em This review is dedicated to the memory of late Bernard Durney, who kindled
my first interest in the theory of the \MC\ many years ago and whose seminal
contributions in this field are often not sufficiently recognized. Peng-Fei Chen
urged me to write this review.  I thank Gopal Hazra, Bidya Karak and Leonid
Kitchatinov for valuable inputs and suggestions on a preliminary version of
the manuscript.} 

\section*{Appendix. Order of magnitude estimates of various terms in the
equation driving the \MC}       

We pointed out in \S3.3 that there should be  pole-equator temperature difference
to give rise to a thermal wind term comparable to the centrifugal term and that
the dissipation term should be negligible compared to these source tersm.  Now
we present some order of magnitude estimates of these terms.

Let us first proceed with the assumption that the dissipation term is negligible
and the thermal wind balance condition (37) holds within the convection zone.
It is easy to argue that the left hand side of (37) is approximately equal to
$$r \sin \theta \, \frac{\pa}{\pa z} \Omega^2 \approx - [\Omega_{\rm eq}^2 - \Omega_{\rm mid}^2], \eqno(A1)$$
where $\Omega_{\rm eq}$ and $\Omega_{\rm mid}$ are the surface values of $\Omega$ at the equator and
at the mid-latitude respectively.  The respective values of frequency at these points are 440 and 400 nHz (see Figure~3), from which the values of $\Omega$ can be obtained by a multiplication with $2 \pi$.  We thus have
$$r \sin \theta \, \frac{\pa}{\pa z} \Omega^2 \approx - [(440)^2 - (400)^2] \times (2 \pi 10^{-9})^2 \; {\rm s}^{-2}
\approx 1.4 \times 10^{-12}  \; {\rm s}^{-2}. \eqno(A2)$$

To make an estimate of the right hand side of (37), we note that the specific entropy of an ideal gas
is given by 
$$S = C_V \ln T  - (\gamma -1) \, C_V \ln \rho + K, $$
where $K$ is a constant. The entropy difference between the equator and the pole at the solar surface (which
is surface of constant $\rho$) is
$$\Delta S = C_V \ln\left( \frac{T_{\rm eq}}{T_{\rm pole}} \right).$$
Taking $\Delta T$ to be the temperature excess of the pole with respect to equator, we have
$$\Delta S \approx - C_V \, \frac{\Delta T}{T_S}, \eqno(A3)$$
where $T_S$ is the temperature of the solar surface and we have made use of the approximation $\ln (1 +x)
\approx x$ for $|x| \ll 1$. Since this entropy difference takes place over an angular separation $\pi/2$,
we have
$$\frac{\pa S}{\pa \theta} \approx - 2 C_V \, \frac{\Delta T}{\pi T_S}, \eqno(A4)$$
Substituting this in the right hand side of (37), we get
$$\frac{1}{r} \frac{g}{\gamma \, C_V}\frac{\pa S}{\pa \theta}
\approx - \frac{2}{ \pi \gamma} \frac{G M_{\odot}}{(0.85 R_{\odot})^3} \frac{\Delta T}{ T_S}, \eqno(A5)$$
where we have taken $r$ to be given by $0.85 R_{\odot}$ corresponding to the middle of the convection zone
and have also used this to calculate $g$.  If we now use the standard values
of solar mass and radius, then we get (taking $\gamma = 1.4$)
$$\frac{1}{r} \frac{g}{\gamma \, C_V} \frac{\pa S}{\pa \theta} \approx 2.8 \times 10^{-7} 
\frac{\Delta T}{T_S} \; {\rm s}^{-2}.
\eqno(A6)$$

Finally, if we equate (A2) and (A6) as required by the thermal wind balance
condition (37), we arrive at
$$\frac{\Delta T}{T_S} \approx 5.0 \times 10^{-6}. \eqno(A7)$$
If we take $T_S$ equal to the temperature 5800 K at the photospheric surface,
then we get a rather low value $\Delta T \approx 2.9 \times 10^{-2}$  K. But,
should we use the photospheric temperature for $T_S$ in (A7)? Choudhuri [153]
has argued that we should use a temperature deeper in the convection zone for
$T_S$ and pointed out that these order of magnitude estimates provide a clue
for understanding the origin of the near-surface shear layer seen in Figure~3.

We now make an order of magnitude estimate of the last term in the equation (27)
of the \MC, the dissipation term, to show that it  would be negligible compared to
the centrifugal term.  If ${\bf K}$ is given by (13), then the last term in (27) is of order
$$\frac{\mu_T |v_m|}{\rho L^3}, $$
where we can take the length scale $L$ to be equal to the thickness $2 \times
10^{10}$ cm of the convection zone.  The quantity $\mu_T/\rho$, called the
kinematic viscosity, is estimated to be about $10^{12}$ within the convection
zone [124]. Taking $|v_m| \approx 10^3$ cm s$^{-1}$, the value of the last term
in (27) comes out to be of order $1.3 \times 10^{-16}$ s$^{-6}$.  Comparing
with (A2), we point out that this term clearly cannot balance the centrifugal term,
which has to be balanced by the thermal wind term, as can be seen in (27).

\def\apj{{Astrophys.\ J.}}
\def\mnras{{Mon.\ Notic.\ Roy.\ Astron.\ Soc.}}
\def\sol{{Solar Phys.}}
\def\aa{{Astron.\ Astrophys.}}
\def\gafd{{\it Geophys.\ Astrophys.\ Fluid Dyn.}}
\def\lrsp{{Living Rev. Solar Phys.}}


\end{document}